\documentclass{aa}
\usepackage{graphicx}

\def\mb#1{\setbox0=\hbox{$#1$}\kern-.025em\copy0\kern-\wd0
\kern-0.05em\copy0\kern-\wd0\kern-.025em\raise.0233em\box0}

\begin{document}
   \title{Gravitational instability of isothermal and polytropic
   spheres }

   \author{P.H. Chavanis}

\institute{ Laboratoire de Physique Quantique, Universit\'e Paul Sabatier, 118
route de Narbonne 31062 Toulouse, France\\
\email{chavanis@irsamc2.ups-tlse.fr}}

   \date{To be included later }

\abstract{We complete previous investigations on the thermodynamics
   of self-gravitating systems by studying the grand canonical, grand
   microcanonical and isobaric ensembles. We also discuss the
   stability of polytropic spheres in connexion with a generalized
   thermodynamical approach proposed by Tsallis. We determine in each
   case the onset of gravitational instability by analytical methods
   and graphical constructions in the Milne plane. We also discuss the
   relation between dynamical and thermodynamical stability of stellar
   systems and gaseous spheres. Our study provides an aesthetic and
   simple approach to this otherwise complicated subject.
  
   }

   \maketitle
%

\section{Introduction}
\label{sec_introduction}

The statistical mechanics of systems interacting via long-range forces
exhibits peculiar features such as negative specific heats,
inequivalence of statistical ensembles and phase transitions. These
curious behaviours have been first discussed in the astrophysical
literature during the elaboration of a thermodynamics for stars
(Eddington 1926), globular clusters (Lynden-Bell \& Wood 1968), black
holes (Hawking 1974) and galaxies (Padmanabhan 1990). They have been
recently rediscovered in different fields of physics such as nuclear
physics, plasma physics, Bose-Einstein condensates, atomic clusters,
two-dimensional turbulence... The main challenge is represented by the
construction of a thermodynamic treatment of systems with long-range
forces and by the analogies and differences among the numerous domains
of application (see Chavanis 2002e and other contributions in that book).

Gravity provides a fundamental example of unshielded long-range
interaction for which ideas of statistical mechanics and
thermodynamics can be developed and tested.  For systems with
long-range interactions, the mean-field approximation is known to be
{\it exact} in a suitable thermodynamic limit. Therefore, the
structure and stability of self-gravitating systems at statistical
equilibrium can be analyzed in terms of the maximization of a
thermodynamical potential. This thermodynamical approach leads to
isothermal configurations which have been studied for a long time in
the context of stellar structure (Chandrasekhar 1942) and galactic
structure (Binney
\& Tremaine 1987). As is well-known, isothermal spheres have infinite
mass so that the system must be confined within a box (Antonov
problem) in order to prevent evaporation and make the thermodynamical
approach rigorous. It is also well-known that isothermal
configurations only correspond to {\it metastable} equilibrium states
(i.e., local maxima of the thermodynamical potential), not true
equilibrium states. These metastable equilibrium states are expected
to be relevant, however, for the timescales contemplated in
astrophysics. In particular, globular clusters described by
Michie-King models are probably in such metastable states.

The series of equilibria of finite isothermal spheres can be
parametrized by the density contrast between the center and the
boundary of the system. For sufficiently low density contrasts, the
system is  thermodynamically stable. However, instability
occurs at sufficiently large concentrations: at some point in
the series of equilibria, the solutions cease to be local maxima of
the thermodynamical potential and become unstable saddle points. The
crucial point to realize is that the onset of instability depends on
the statistical ensemble considered: microcanonical (MCE), canonical
(CE) or grand canonical (GCE). This contrasts with ordinary systems,
with short range interactions, for which the statistical ensembles are
equivalent at the thermodynamic limit, except near a phase
transition. This suggests that gravitating systems in virial
equilibrium are similar to normal (extensive) systems at the verge of
a phase transition (Padmanabhan 1990).

The thermodynamical stability of self-gravitating systems can be
studied by different technics. Horwitz \&
Katz (1978) use a field theory and write the density of states,
the partition function and the grand partition function in MCE, CE and
GCE respectively as a path integral for a formal field $\phi$. In the
mean-field approximation, the integral is dominated by the distribution
$\phi_{0}$ which {maximizes} a specific action $A\lbrack
\phi\rbrack$. In the grand canonical ensemble, $A_{GCE}\lbrack
\phi\rbrack$ is the Liouville action. Horwitz \& Katz (1978) solve the problem 
numerically and find that the series of equilibria become unstable for
a density contrast $1.58$, $32.1$ and $709$ in GCE, CE and MCE
respectively. It is normal that the critical density contrast (hence
the stability of the system) increases when more and more constraints
are added on the system (conservation of mass in CE, conservation of
mass and energy in MCE). This field theory has been rediscussed
recently by de Vega \& Sanchez (2002) who confirmed previous results
and proposed interesting developements.

The thermodynamical stability of self-gravitating systems can also be
settled by studying whether an isothermal sphere is a maximum or a
saddle point of an appropriate thermodynamical potential: the entropy
in MCE, the free energy in CE and the grand potential in GCE. The
change of stability can be determined very easily from the topology of
the equilibrium phase diagram by using the turning point criterion of
Katz (1978) who has extended Poincar\'e's theory on linear series of
equilibria (see also Lynden-Bell \& Wood 1968). This method is very
powerful but it does not provide the form of the perturbation profile
that triggers the instability. This perturbation profile can be
obtained by computing explicitly the second order variations of the
thermodynamical potential and reducing the problem of stability to the
study of an eigenvalue equation. This study was first performed by
Antonov (1962) in MCE and revisited by Padmanabhan (1989) with a
simpler mathematical treatment.  This analysis was extended in CE by
Chavanis (2002a) who showed in addition the equivalence
between thermodynamical stability and dynamical stability with respect
to Navier-Stokes equations (Jeans problem). Remarkably, this stability
analysis can be performed analytically by using simple graphical
constructions in the Milne plane.

In the present paper, we propose to extend these analytical methods to
more general situations in order to provide a complete description of
the thermodynamics of spherical self-gravitating systems. In
Sec. \ref{sec_thermo}, we review the stability limits of isothermal
spheres in different ensembles by using the turning point
criterion. In Sec. \ref{sec_stab}, we consider specifically the grand
canonical and grand microcanonical ensembles and evaluate the second
order variations of the associated thermodynamical potential.  In
Sec. \ref{sec_field}, we briefly discuss the connexion between
thermodynamics and statistical mechanics (and field theory) for
self-gravitating systems. In Sec. \ref{sec_tsallis}, we consider the
case of generalized thermodynamics proposed by Tsallis (1988) and
leading to stellar polytropes (Plastino \& Plastino 1997, Taruya \&
Sakagami 2002a,b, Chavanis 2002b). In Sec. \ref{sec_isobaric}, we
discuss the stability of isothermal and polytropic spheres under an
external pressure (Bonnor problem). We provide a new and entirely
analytical solution of this old problem. Finally, in Sec. \ref{sec_st}
we discuss the relation between dynamical and thermodynamical
stability of stellar systems and gaseous spheres.

\section{Thermodynamics of self-gravitating systems}
\label{sec_thermo}

\subsection{Thermodynamical ensembles}
\label{sec_ensembles}

Consider a system of $N$ particles, each of mass $m$, interacting via Newtonian gravity. Let $f({\bf r},{\bf v},t)$ denote the distribution function of the system and $\rho({\bf r},t)=\int fd^{3}{\bf v}$ the spatial density.  The total mass and total energy can be expressed as
\begin{equation}
M=\int \rho d^{3}{\bf r},
\label{ens1}
\end{equation} 
\begin{equation}
E={1\over 2}\int f v^{2}d^{3}{\bf r}d^{3}{\bf v}+{1\over 2}\int \rho\Phi d^{3}{\bf r}=K+W,
\label{ens2}
\end{equation}
where $K$ is the kinetic energy and $W$ the potential energy.  The
gravitational potential $\Phi$ is related to the spatial density by
the Poisson equation
\begin{equation}
\Delta\Phi=4\pi G\rho.
\label{ens3}
\end{equation}
The Boltzmann entropy is given by the standard formula (within an
additional constant term)
\begin{equation}
S=-k\int \biggl\lbrace {f\over m}\ln {f\over m}-{f\over m}\biggr\rbrace d^{3}{\bf r}d^{3}{\bf v},
\label{ens4}
\end{equation}
which can be obtained by a combinatorial analysis (Ogorodnikov 1965).
For the moment, we shall assume that the system is confined within a
spherical box of radius $R$ so that its volume is fixed.

In the {\it microcanonical ensemble}, the system is isolated and
conserves energy $E$ and particle number $N$ (or mass
$M=Nm$). Statistical equilibrium corresponds to the state that
maximizes the entropy $S$ at fixed $E$ and $N$. Introducing Lagrange
multipliers $1/T$ and $-\mu/T$ for each constraint, the first order
variations of entropy satisfy the relation
\begin{equation}
\delta S-{1\over T}\delta E+{\mu\over T}\delta N=0,
\label{ens5}
\end{equation}
where $T$ is the temperature and $\mu$ the chemical potential. Eq. (\ref{ens5})
can be regarded as the first principle of thermodynamics.

In the {\it canonical ensemble}, the temperature and the particle
number are fixed allowing the energy to fluctuate. In that case, the
relevant thermodynamical parameter is the free energy (more precisely
the Massieu function) $J=S-{1\over T}E$ which is related to $S$ by a
Legendre transformation. According to Eq. (\ref{ens5}), the first
variations of $J$ satisfy $\delta J=-E\delta ({1\over T})-{\mu\over
T}\delta N$.  Therefore, at statistical equilibrium, the system is in the
state that maximizes the free energy $J$ at fixed particle number
$N$ and temperature $T$. 

In the {\it grand canonical ensemble}, both temperature and chemical
potential are fixed, allowing the energy and the particle number to
fluctuate. The relevant thermodynamical potential is now the grand
potential $G=S-{1\over T}E+{\mu\over T}N$. Its first variations
satisfy $\delta G=-E\delta ({1\over T})+N\delta ({\mu\over T} )$. At
statistical equilibrium, the system is in the state that maximizes
$G$ at fixed $T$ and $\mu/T$.

The microcanonical, canonical and grand canonical ensembles are the
most popular. However, we are free to define other thermodynamical
ensembles. For example, Lecar \& Katz (1981) have introduced a {\it
grand microcanonical ensemble} (GMCE) in which the fugacity
and the energy are fixed. This corresponds to a thermodynamical
potential ${\cal K}=S+{\mu\over T}N$ whose first variations satisfy
$\delta {\cal K}={1\over T}\delta E+N\delta ({\mu\over T} )$. At
statistical equilibrium, the system is in the state that maximizes
${\cal K}$ at fixed $E$ and $\mu/T$.

\subsection{The isothermal distribution}
\label{sec_iso}

For extensive systems, it is well-known that the statistical ensembles
described previously are all equivalent at the thermodynamic
limit. Therefore, the choice of a particular ensemble (in general the
grand canonical one) is only dictated by reasons of convenience. This
is {\it not} the case for systems with long-range interactions such as
gravitational systems. Indeed, the statistical ensembles are not
interchangeable and the stability limits differ from one ensemble to
the other. Therefore, the choice of the ensemble is imposed by the
physical properties of the system under consideration. If we want to
describe in terms of statistical mechanics globular clusters
(Lynden-Bell \& Wood 1968) and elliptical galaxies (Lynden-Bell 1967,
Hjorth \& Madsen 1993, Chavanis \& Sommeria 1998), which can be
treated as isolated systems in a first approximation, the good choice
is the microcanonical ensemble. If now the system is in contact with a
radiation background imposing its temperature, the relevant ensemble
is the canonical one. This ensemble may be appropriate to star or
galaxy formation (Penston 1969, Chavanis 2002a,b), white dwarf and
neutron stars (Hertel \& Thirring 1971, Chavanis 2002c,d) and dark
matter made of massive neutrinos (Bilic \& Viollier 1997, Chavanis
2002g). This canonical description is also exact in a model
of self-gravitating Brownian particles (Chavanis et al. 2002, Sire \&
Chavanis 2002). Finally, the grand canonical ensemble may be
appropriate to the interstellar medium (or possibly the large-scale
structures of the universe), assuming that a statistical description
is relevant (de Vega et al. 1998). Possible applications of the grand
microcanonical ensemble are discussed by Lecar \& Katz (1981).

If we just cancel the first order variations of the thermodynamical
potential (under constraints appropriate to the ensemble considered),
it is straightforward to check that {\it each} ensemble yields the
Maxwell-Boltzmann distribution
\begin{equation}
f=m \ e^{\mu/kT}\ e^{-{m\over kT}({v^{2}\over 2}+\Phi)}.
\label{iso1}
\end{equation} 
Therefore, the statistical equilibrium state of a self-gravitating system
is {\it isothermal} whatever the ensemble
considered. The differences will come from the second order variations
of the thermodynamical potential. The choice of the
statistical ensemble will affect the {\it stability} of the system.

The density field associated with the distribution function (\ref{iso1}) is given by
\begin{equation}
\rho=\biggl ({2\pi \over \beta}\biggr )^{3/2}ze^{-{\beta\Phi}},
\label{iso2}
\end{equation} 
where we have defined the fugacity $z$ and the inverse temperature $\beta$ by 
the relations 
\begin{equation}
z=m e^{\mu/kT},\qquad \beta={m\over kT}.
\label{iso3}
\end{equation} 
The distribution function (\ref{iso1}) can then  be written in
terms of $\rho$ as
\begin{equation}
f=\biggl ({m\over 2\pi kT}\biggr )^{3/2}\rho({\bf r})e^{-{mv^{2}\over 2kT}}.
\label{iso4}
\end{equation} 
This distribution function is a {\it global} maximum of the
thermodynamical potential under the usual constraints {\it and} for a
given density field $\rho({\bf r})$ (see Padmanabhan 1989, Chavanis
2002a).  Substituting this optimal distribution in Eqs. (\ref{ens2})
and (\ref{ens4}), we get
\begin{equation}
E={3\over 2}NkT+{1\over 2}\int\rho\Phi d^{3}{\bf r},
\label{iso5}
\end{equation} 
\begin{equation}
{S\over k}={5N\over 2}+{3N\over 2}\ln\biggl ({2\pi kT\over m}\biggr )-\int{\rho\over m}\ln{\rho\over m}d^{3}{\bf r}.
\label{iso6}
\end{equation} 
It will be more convenient in the sequel to study the stability problem directly from  these equations, without loss of generality. 

\subsection{The Emden equation and the Milne variables}
\label{sec_emden}

Inserting the relation (\ref{iso2}) in Eq. (\ref{ens3}), we find
that the gravitational potential $\Phi$ is a solution of the Boltzmann-Poisson
equation
\begin{equation}
\Delta\Phi=4\pi GAe^{-\beta\Phi}.
\label{emden0}
\end{equation} 
If we introduce the function $\psi=\beta(\Phi-\Phi_{0})$
where $\Phi_{0}$ is the gravitational potential at $r=0$, the density
field can be written
\begin{equation}
\rho=\rho_{0}e^{-\psi},
\label{emden1}
\end{equation} 
where $\rho_{0}$ is the central density. Introducing the normalized
distance $\xi=(4\pi G\beta\rho_{0})^{1/2}r$, the Boltzmann-Poisson
equation can be written in the standard Emden form (Chandrasekhar
1942)
\begin{equation}
{1\over\xi^{2}}{d\over d\xi}\biggl (\xi^{2}{d\psi\over d\xi}\biggr )=e^{-\psi},
\label{emden2}
\end{equation} 
with $\psi=\psi'=0$ at $\xi=0$. This equation has to be solved between $\xi=0$ and $\xi=\alpha$ corresponding to the normalized box radius
\begin{equation}
\alpha=(4\pi G\beta\rho_{0})^{1/2}R.
\label{alp}
\end{equation} 

We also recall the definition of the Milne variables that will be used in the sequel
\begin{equation}
u={\xi e^{-\psi}\over \psi'},\qquad v=\xi\psi'.
\label{emden3}
\end{equation}
These variables satisfy the identities
\begin{equation}
{1\over u}{du\over d\xi}={1\over \xi}(3-v-u),
\label{emden4}
\end{equation} 
\begin{equation}
 {1\over v}{dv\over d\xi}={1\over \xi}(u-1),
\label{emden5}
\end{equation} 
which can be directly derived from Eq. (\ref{emden2}).

The phase portrait of classical isothermal spheres is represented in Fig. \ref{valcrit} and forms a spiral. The spiral starts at $(u,v)=(3,0)$ for $\xi=0$ and tends to the limit point $(u_{s},v_{s})=(1,2)$ as $\xi\rightarrow +\infty$. This limit point corresponds to the singular solution $e^{-\psi_{s}}=2/\xi^{2}$. For finite isothermal spheres, we must consider only the part of the curve between $\xi=0$ and $\xi=\alpha$.

\subsection{The equilibrium phase diagram}
\label{sec_eq}

The thermodynamical parameters can be expressed in terms of the values
of the Milne variables at the normalized box radius $\alpha$. Writing
$u_{0}=u(\alpha)$ and $v_{0}=v(\alpha)$, the energy and the
temperature can be written as (Padmanabhan 1989, Chavanis 2002a):
\begin{equation}
\Lambda\equiv -{ER\over GM^{2}}={1\over v_{0}}\biggl ({3\over 2}-u_{0}\biggr ),
\label{eq1}
\end{equation} 
\begin{equation}
\eta\equiv {\beta GM\over R}=v_{0}.
\label{eq2}
\end{equation} 

We need also to express  $\alpha$ in terms of the fugacity $z$. Using the expression of the gravitational potential at the box radius $\Phi(R)=-GM/R$ and comparing Eqs. (\ref{iso2}) and (\ref{emden1}), we get
\begin{equation}
\biggl ({2\pi \over \beta}\biggr )^{3/2}ze^{\eta}=\rho_{0}e^{-\psi(\alpha)},
\label{eq3}
\end{equation} 
Expressing $\rho_{0}$ in terms of $\alpha$ by Eq. (\ref{alp}) and
introducing the Milne variables (\ref{emden3}), we obtain
\begin{equation}
\chi\equiv {4(2\pi)^{5}G^{2}R^{4}z^{2}\over\beta}=u_{0}^{2}
v_{0}^{2}e^{-2v_{0}}.
\label{eq4}
\end{equation}
We can also normalize the fugacity by the energy instead of the
temperature. Combining Eqs. (\ref{eq4}), (\ref{eq2}) and
(\ref{eq1}), we get
\begin{equation}
\nu\equiv -16(2\pi)^{10}G^{5}R^{7}z^{4}E=u_{0}^{4}v_{0}^{5}e^{-4v_{0}}\biggl ({3\over 2}-u_{0}\biggr ).
\label{eq5}
\end{equation}

\begin{figure}
\centering
\includegraphics[width=8.5cm]{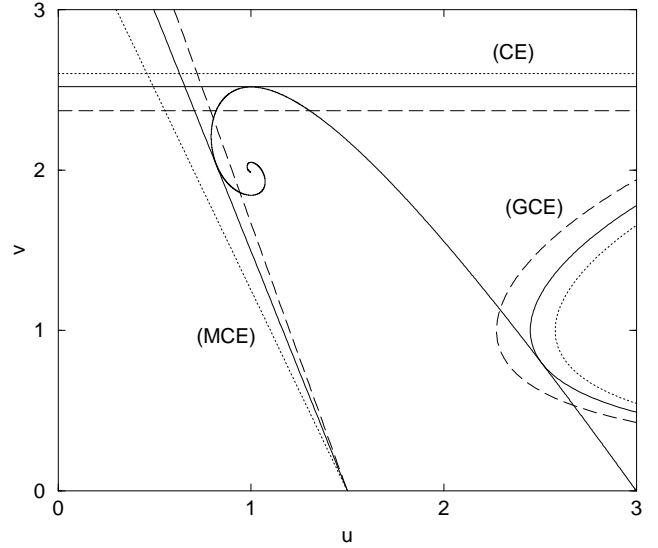}
\caption{This graphical construction, first introduced 
by Padmanabhan (1989) in MCE, shows very simply the existence of an
upper bound for $\Lambda$, $\eta$ and $\chi$ above which there is no
hydrostatic equilibrium for an isothermal gas (dotted line:
$\Lambda>\Lambda_{c}$; solid line: $\Lambda=\Lambda_{c}$; dashed line:
$\Lambda<\Lambda_{c}$ with similar convention for $\eta$ and $\chi$).}
\label{valcrit}
\end{figure}

It is easy to show that statistical equilibrium states only exist for
sufficiently small values of the control parameters $\Lambda$, $\eta$,
$\chi$ and $\nu$ defined previously. Above a critical value, there is
no possible equilibrium state and the system is expected to
collapse. For example, for $\Lambda\ge \Lambda_{c}=0.335$ (which
corresponds to sufficiently negative energies), we have the well-know
gravothermal catastrophe (Antonov 1962, Lynden-Bell \& Wood
1968). Similarly, for $\eta\ge\eta_{c}=2.52$ (which corresponds to
small temperatures or large masses), we have an isothermal collapse
(Chavanis 2002a, Chavanis et al. 2002). We can show the existence of
such bounds by a simple graphical construction, without numerical
work. Indeed, the equilibrium structure of the system (characterized
by its degree of central concentration $\alpha$) is determined by the
intersection between the spiral in the $(u,v)$ plane and the curves
defined by Eqs. (\ref{eq1}), (\ref{eq2}) and (\ref{eq4}). If there is
no intersection, the system cannot be in statistical equilibrium. This
graphical construction is made explicitly in Fig. \ref{valcrit} in
MCE, CE and GCE.

\begin{figure}
\centering
\includegraphics[width=8.cm]{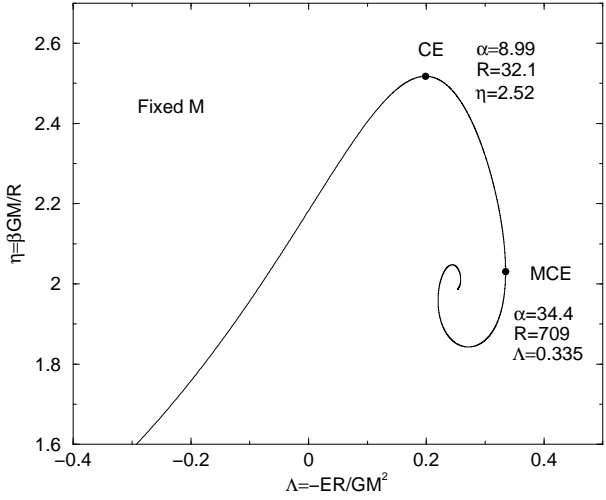}
\caption{Temperature vs energy plot at fixed particle number. The series of equilibria becomes unstable at MCE in the microcanonical ensemble and at CE in the canonical ensemble.}
\label{Leta}
\end{figure}

\begin{figure}
\centering
\includegraphics[width=8.8cm]{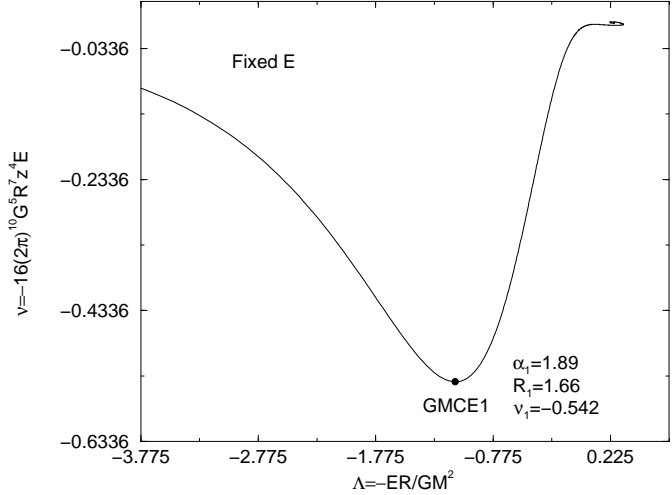}
\caption{Fugacity vs particle number at fixed energy. The series of equilibria becomes unstable at GMCE1 in the grand microcanonical ensemble. }
\label{Lnu}
\end{figure}

\begin{figure}
\centering
\includegraphics[width=8.8cm]{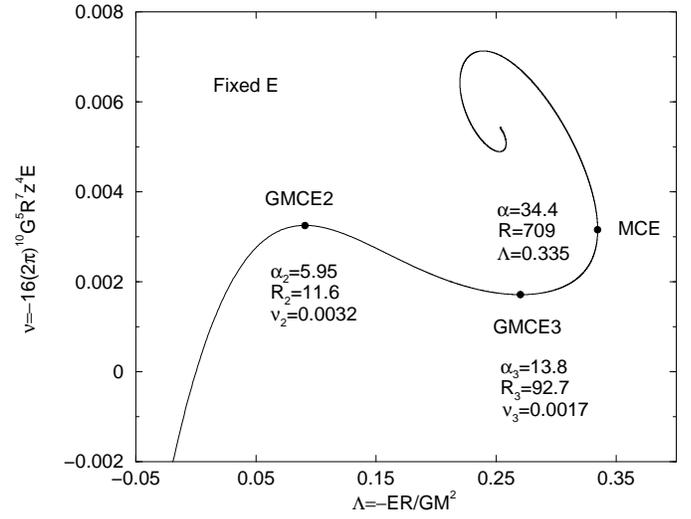}
\caption{Enlargement of Fig. \ref{Lnu} near the spiral. In the grand microcanonical ensemble, a change of stability occurs at GMCE2 and GMCE3. Between these points the series of equilibria is stable again. After GMCE3, the series becomes and remains unstable. }
\label{LnuZOOM}
\end{figure}

\begin{figure}
\centering
\includegraphics[width=8.5cm]{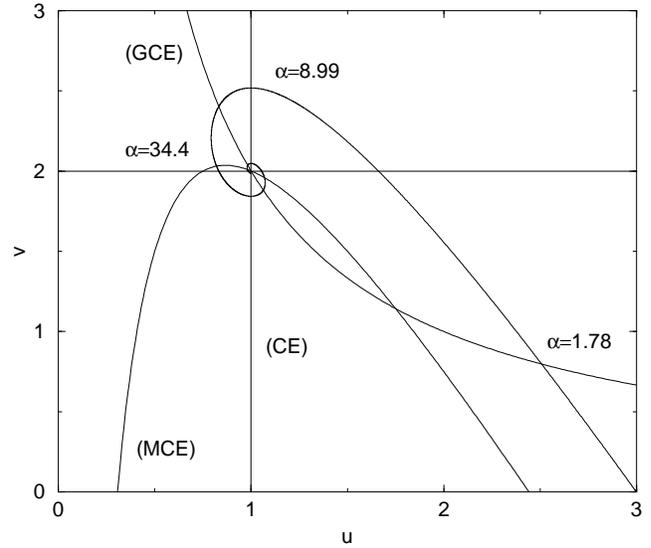}
\caption{Graphical construction determining the turning points of the control parameter at which a change of stability occurs in MCE, CE and GCE.}
\label{uvcrit1}
\end{figure}

In the microcanonical ensemble, the control parameters are the energy
$E$ and the particle number $N$. The parameter conjugate to the energy
(with respect to the entropy $S$) is the inverse temperature
$\beta$. The curve giving the inverse temperature as a function of
energy for a fixed particle number $N$ is drawn in
Fig. \ref{Leta}. The parameter conjugate to the particle number $N$ is
the ratio $-\mu/T$, related to the fugacity $z$. In order to represent
the fugacity as a function of $N$, for a fixed energy, we use
Eqs. (\ref{eq1}) and (\ref{eq5}). The $-\mu/T$ vs $N$ curve can then
be deduced from Figs. \ref{Lnu} and \ref{LnuZOOM}. According to
standard turning point arguments (Katz 1978), the series of equilibria
becomes unstable at the point of minimum energy (for a given mass) or
at the point of minimum mass (for a given negative energy). The condition
$d\Lambda/d\alpha=0$ implies (Padmanabhan 1989)
\begin{equation}
4u_{0}^{2}+2u_{0}v_{0}-11u_{0}+3=0.
\label{eq6}
\end{equation} 
The values of $\alpha$ at which the parameter $\Lambda$ is extremum are determined by the intersections between the spiral in the $(u,v)$ plane and the parabole defined by Eq. (\ref{eq6}), see Fig. \ref{uvcrit1}. The change of stability occurs for the smallest value of $\alpha$. Its numerical value is $\alpha=34.4$ corresponding to a density contrast of $709$. Gravitational instability occurs in MCE when the specific heats $C_{V}=(\partial E/\partial T)_{N,V}$ becomes zero, passing from negative to {\it positive} values.

\begin{figure}
\centering
\includegraphics[width=8.8cm]{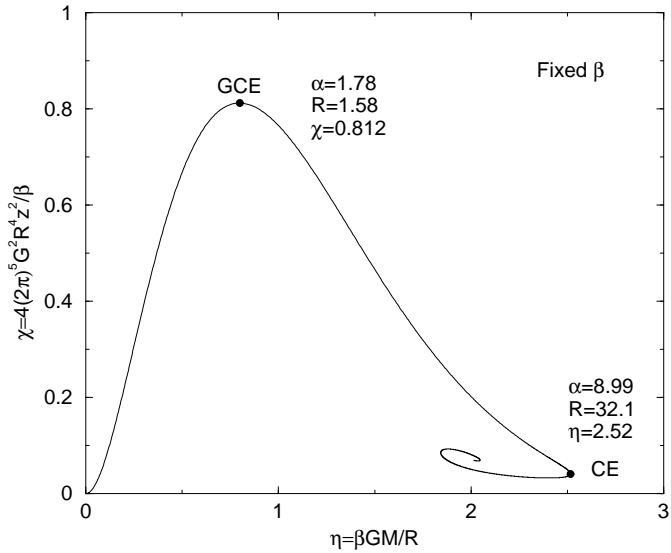}
\caption{Fugacity vs mass at  fixed temperature. The series of equilibria becomes unstable at GCE in the grand canonical ensemble.}
\label{etachi}
\end{figure}

\begin{figure}
\centering
\includegraphics[width=8.8cm]{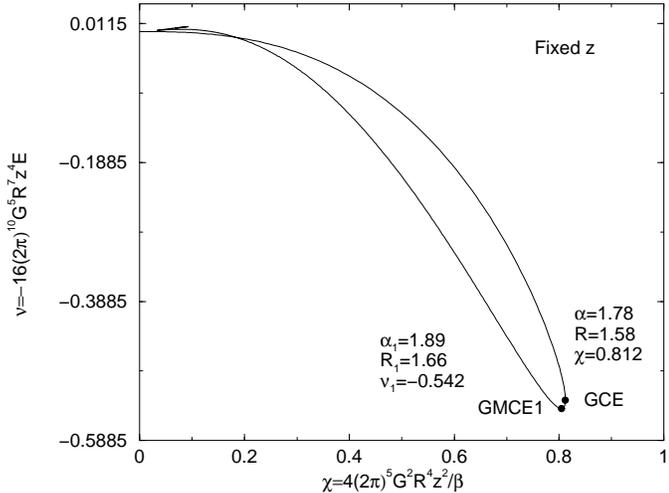}
\caption{Energy vs temperature at fixed fugacity.}
\label{chinu}
\end{figure}

\begin{figure}
\centering
\includegraphics[width=8.8cm]{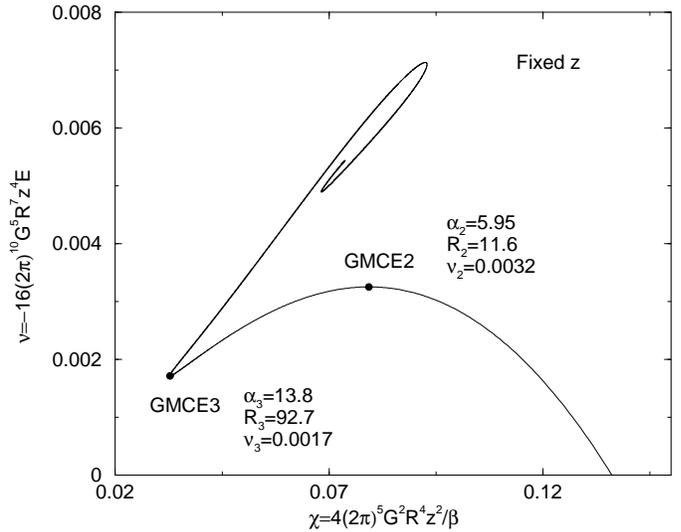}
\caption{Enlargement of Fig. \ref{chinu} near the spiral. }
\label{chinuZOOM}
\end{figure}

\begin{figure}
\centering
\includegraphics[width=8.8cm]{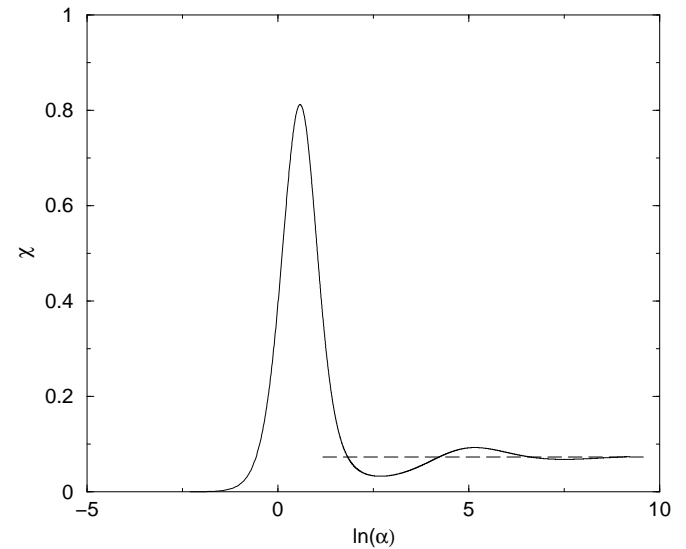}
\caption{Evolution of $\chi$ as a function of the central concentration $\alpha$. In dashed line we have represented the asymptotic value $\chi_{s}=4e^{-4}$ corresponding to the singular solution. The  $\Lambda(\alpha)$ and $\eta(\alpha)$ curves  have a similar behaviour. }
\label{alphachi}
\end{figure}

\begin{figure}
\centering
\includegraphics[width=8.8cm]{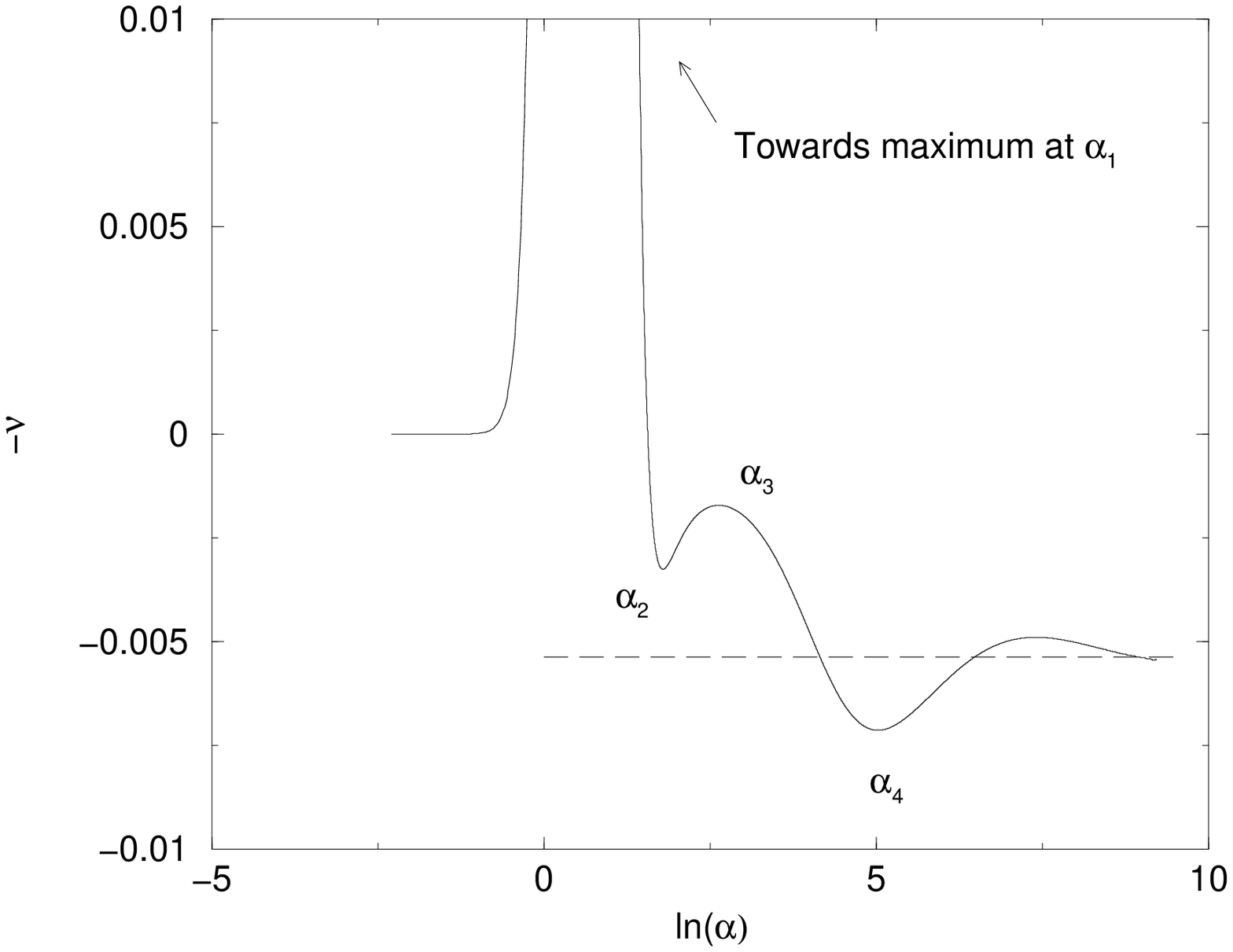}
\caption{Evolution of $\nu$ as a function of the central concentration $\alpha$. In dashed line we have represented the asymptotic value $\nu_{s}=2^{4}e^{-8}$ corresponding to the singular solution. Contrary to Fig. \ref{alphachi}, the second extremum $\alpha_{2}$ is {\it above} the asymptote $\nu=\nu_{s}$. This appears to be the reason for the regain of stability between $\alpha_{2}$ and $\alpha_{3}$.}
\label{alphanuZOOM}
\end{figure}

In the canonical ensemble, the control parameters are the inverse
temperature $\beta$ and the particle number $N$. The parameter
conjugate to the inverse temperature (with respect to the free energy
$J$) is $-E$. The curve giving $-E$ as a function of $\beta$ for a
fixed particle number can be deduced from Fig. \ref{Leta}. The
parameter conjugate to the particle number is $-\mu/T$, related to the
fugacity $z$. We can represent the fugacity as a function of mass, for
a given temperature, by using the relations (\ref{eq4}) and
(\ref{eq2}). The corresponding curve can be deduced from the $\eta$ vs
$\chi$ plot represented in Fig. \ref{etachi}. In the canonical
ensemble, the series of equilibria becomes unstable at the point of
minimum temperature (for a given mass) or at the point of maximum mass
(for a given temperature). The condition $d\eta/d\alpha=0$ is
equivalent to (Chavanis 2002a)
\begin{equation}
u_{0}=1.
\label{eq7}
\end{equation} 
The values of $\alpha$ at which the parameter $\eta$ is extremum are
determined by the intersections between the spiral in the $(u,v)$ plane
and the straight line defined by Eq. (\ref{eq7}), see
Fig. \ref{uvcrit1}. The change of stability occurs for the smallest
value of $\alpha$. Its numerical value is $\alpha=8.99$ corresponding
to a density contrast of $32.1$. Gravitational instability in CE
occurs when the specific heats becomes {\it negative} passing by
$C_{V}=\infty$.

In the grand canonical ensemble, the control parameters are the
inverse temperature $\beta$ and the ratio $\mu/T$ related to the fugacity
$z$. The parameter conjugate to the inverse temperature (with respect to the grand potential $G$) is $-E$. The
curve giving $-E$ as a function of $\beta$ for a fixed fugacity $z$ is
determined by Eqs. (\ref{eq4}) and (\ref{eq5}) and can be deduced from
Figs. \ref{chinu} and
\ref{chinuZOOM}. The parameter conjugate to the ratio
$\mu/T$ is the particle number $N$. The curve giving $N$ as a function
of $z$ for a given temperature can be deduced from
Fig. \ref{etachi}.  In the grand canonical ensemble,
the series of equilibria becomes unstable at the point of maximum
temperature (for a given fugacity) or at the point of maximum fugacity
(for a given temperature). The condition $d\chi/d\alpha=0$ implies
\begin{equation}
u_{0}v_{0}=2.
\label{eq8}
\end{equation} 
The values of $\alpha$ at which the parameter $\chi$ is extremum (see
Fig. \ref{alphachi}) are determined by the intersections between the
spiral in the $(u,v)$ plane and the hyperbole defined by
Eq. (\ref{eq8}), see Fig. \ref{uvcrit1}. The change of stability
occurs for the smallest value of $\alpha$. Its numerical value is
$\alpha=1.78$ corresponding to a density contrast of $1.58$.

\begin{figure}
\centering
\includegraphics[width=8.5cm]{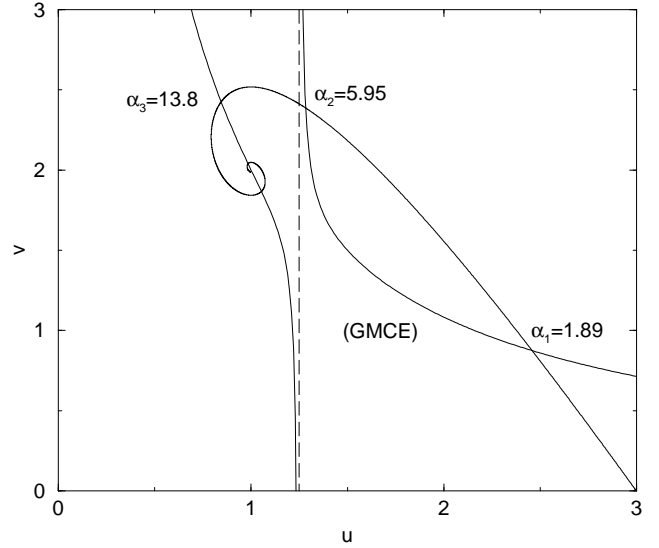}
\caption{Graphical construction determining the turning points of the control parameter $\nu$ in  GMCE. The series of equilibria is stable for $\alpha\le\alpha_{1}$, unstable for $\alpha_{1}\le \alpha\le\alpha_{2}$, stable again for $\alpha_{2}\le \alpha\le\alpha_{3}$ and unstable for 
$\alpha\ge\alpha_{3}$.}
\label{uvcrit2}
\end{figure}

In the grand microcanonical ensemble, the control parameters are the
energy $E$ and the ratio $\mu/T$ related to the fugacity $z$. The
parameter conjugate to the energy (with respect to the potential ${\cal K}$)
is $\beta$. The curve giving $\beta$ as a function of $E$ for a fixed
fugacity $z$ can be deduced from Figs. \ref{chinu} and
\ref{chinuZOOM}.  The parameter conjugate to the ratio $\mu/T$ is the
particle number $N$. The $N$ vs $\mu/T$ plot for a given energy can be
deduced from Figs. \ref{Lnu} and \ref{LnuZOOM}.  In the grand
microcanonical ensemble, the series of equilibria becomes unstable at
the point of maximum energy (for a given fugacity) or at the point of
maximum fugacity (for a given positive energy). The condition $d\nu/d\alpha=0$
implies
\begin{equation}
8u_{0}^{2}v_{0}-10u_{0}v_{0}-17u_{0}+21=0.
\label{eq9}
\end{equation} 
The values of $\alpha$ at which the parameter $\nu$ is extremum (see
Fig. \ref{alphanuZOOM}) are determined by the intersections between
the spiral in the $(u,v)$ plane and the curve defined by
Eq. (\ref{eq9}), see Fig. \ref{uvcrit2}. The series of equilibria
becomes unstable for $\alpha_{1}=1.89$ (density contrast $1.66$) which
corresponds to the first intersection. At that point, the $\nu$ vs
$\chi$ curve rotates clockwise (see Fig. \ref{chinu}). However, at the
second turning point $\alpha_{2}=5.95$ (density contrast $11.6$) the
$\nu$ vs $\chi$ curve rotates anticlockwise (see Fig. \ref{chinuZOOM})
so that stability is regained (this is proved in Katz 1978). Stability
is lost again at $\alpha_{3}=13.8$ (density contrast $92.7$) and never
recovered afterwards. This curious behaviour in GMCE was first noted
by Lecar \& Katz (1981). A similar lost-gain-lost stability behaviour
also occurs for self-gravitating fermions and hard sphere models in
MCE and CE (Chavanis \& Sommeria 1998, Chavanis 2002c). In that case,
it is related to the occurence of phase transitions between
``gaseous'' and ``condensed'' states.

\section{Thermodynamical stability of self-gravitating systems}
\label{sec_stab}

We now determine the condition of thermodynamical stability by
computing the second order variations of the thermodynamical potential
explicitly. This study has already been done in MCE (Padmanabhan 1989) 
and CE (Chavanis 2002a).  In the following, we perform the
analysis in GCE and GMCE.

\subsection{The grand canonical ensemble}
\label{sec_gc}

The second order variations of the grand potential $G$ are 
\begin{equation}
\delta^{2}G=-{k\over m}\int {(\delta\rho)^{2}\over 2\rho}d^{3}{\bf r}-{1\over 2T}\int\delta\rho\delta\Phi d^{3}{\bf r}.
\label{gc1}
\end{equation} 
An isothermal sphere is stable in the grand canonical ensemble if
$\delta^{2}G\le 0$ for all variations $\delta\rho$ of the density
profile. We shall restrict ourselves to spherically symmetric
perturbations since only such perturbations can trigger gravitational
instability for non rotating systems (Horwitz \& Katz 1978). In the
grand canonical ensemble, the mass of the system is not
conserved. Therefore, the potential at the wall changes according to
\begin{equation}
\delta\Phi(R)=-{G\delta M\over R}.
\label{gc2}
\end{equation} 
On the other hand, according to the Gauss theorem, we have
\begin{equation}
{d\delta\Phi\over dr}(R)={G\delta M\over R^{2}}.
\label{gc3}
\end{equation} 
These boundary conditions suggest the introduction of the new function 
\begin{equation}
X(r)=r\delta\Phi(r),
\label{gc4}
\end{equation}
which satisfies
\begin{equation}
X(0)=0,\qquad X''(0)=0,\qquad X'(R)=0,
\label{gc5}
\end{equation}
where we have used Eqs. (\ref{gc2}) and (\ref{gc3}) to obtain the
third equality (the second equality is obvious for a spherically
symmetric system). 

\begin{figure}
\centering
\includegraphics[width=8.5cm]{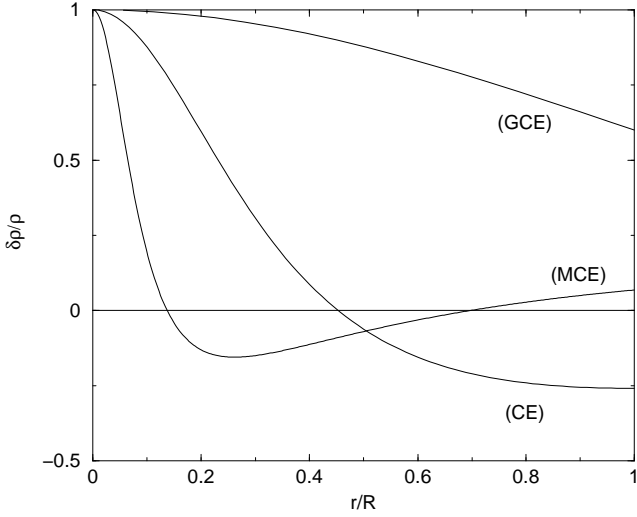}
\caption{First mode of instability in MCE, CE and GCE.}
\label{profilpert}
\end{figure}

The variations $\delta\Phi$ and $\delta\rho$ are related to each other by the Poisson equation
\begin{equation}
{1\over r^{2}}{d\over dr}\biggl (r^{2}{d\delta\Phi\over dr}\biggr )=4\pi G\delta\rho.
\label{gc6}
\end{equation}
In terms of the new variable (\ref{gc4}), Eq. (\ref{gc6}) becomes
\begin{equation}
{X''\over r}=4\pi G\delta\rho.
\label{gc7}
\end{equation}
Expressing the second order variations of $G$ in terms of $X$, we get
\begin{equation}
\delta^{2}G=-{k\over m}{1\over 8\pi G^{2}}\int_{0}^{R}{(X'')^{2}\over\rho}dr-{1\over 2TG}\int_{0}^{R}X''Xdr.
\label{gc8}
\end{equation}
Integrating by parts, we can put Eq. (\ref{gc8}) in the form 
\begin{equation}
\delta^{2}G={1\over 2G^{2}}\int_{0}^{R}dr X'\biggl\lbrack {G\over T}+{k\over m}{d\over dr}\biggl ({1\over 4\pi \rho}{d\over dr}\biggr )\biggr\rbrack X'.
\label{gc9}
\end{equation}
The boundary terms arising from the integration by parts are seen to vanish by virtue of Eq. (\ref{gc5}). We can therefore reduce the stability analysis of isothermal spheres in the grand canonical ensemble to the eigenvalue problem
\begin{equation}
\biggl\lbrack {k\over m}{d\over dr}\biggl ({1\over 4\pi \rho}{d\over dr}\biggr )+{G\over T}\biggr\rbrack X'_{\lambda}(r)=\lambda X'_{\lambda}(r).
\label{gc10}
\end{equation}
If all the eigenvalues $\lambda$ are negative, then the isothermal
sphere is a maximum of the thermodynamical potential $G$ (and is
therefore stable in GCE). If at least one eigenvalue is positive, it
is an unstable saddle point.  The condition of marginal stability
$\lambda=0$ determines the transition between stability and
instability. We thus have to solve the equation
\begin{equation}
{k\over m}{d\over dr}\biggl ({1\over 4\pi \rho}{dX'\over dr}\biggr )+{GX'\over T}=0.
\label{gc11}
\end{equation}
Integrating once and using the boundary conditions (\ref{gc5}) at $r=0$, the constant of integration is seen to vanish and we get
\begin{equation}
{k\over m} {1\over 4\pi \rho}X''+{GX\over T}=0.
\label{gc12}
\end{equation}
Introducing the dimensionless variables defined in Sec. \ref{sec_emden}, we are led to solve the problem  
\begin{equation}
e^{\psi} X''+X=0,
\label{gc13}
\end{equation}
\begin{equation}
X(0)=X'(\alpha)=0.
\label{gc14}
\end{equation}
Let us denote by
\begin{equation}
{\cal L}\equiv e^{\psi}{d^{2}\over d\xi^{2}}+1,
\label{lb1}
\end{equation} 
the differential operator which occurs in Eq. (\ref{gc13}).  Using the
Emden equation (\ref{emden2}), we readily check that
\begin{equation}
{\cal L}(\xi)=\xi,\qquad {\cal L}(\xi^{2}\psi')=2\xi,\qquad {\cal L}(\xi\psi)=\xi+\xi\psi.
\label{lb2}
\end{equation}
Therefore, the general solution of Eq. (\ref{gc13}) which satisfies the condition $X=0$ at the origin is
\begin{equation}
X(\xi)=c_{1}\xi(\xi \psi'-2),
\label{gc15}
\end{equation}
where $c_{1}$ is an arbitrary constant.  The boundary condition
$X'(\alpha)=0$ implies
\begin{equation}
\alpha^{2}e^{-\psi(\alpha)}-2=0.
\label{gc16}
\end{equation}
This relation determines the value of $\alpha$ at which the series of equilibria becomes unstable in the grand canonical ensemble. In terms of the Milne variables, Eq. (\ref{gc16}) is equivalent to
\begin{equation}
u_{0}v_{0}=2,
\label{gc17}
\end{equation}
which is precisely the condition (\ref{eq8}) obtained with the turning point criterion. We are now in a position to determine the structure of the perturbation profile that triggers the instability in GCE. Using Eq. (\ref{gc7}), we have
\begin{equation}
{\delta\rho\over\rho_{0}}={X''\over 4\pi \xi},
\label{q1}
\end{equation} 
with Eq.  (\ref{gc15}) for $X$. This yields
\begin{equation}
{\delta\rho\over \rho_{0}}={c_{1}\over 4\pi}e^{-\psi}(2-\xi\psi'),
\label{gc18}
\end{equation}  
or, introducing the Milne variables,
\begin{equation}
{\delta\rho\over \rho}={c_{1}\over 4\pi}(2-v).
\label{gc19}
\end{equation} 
We note that Eq. (\ref{gc19}) coincides with the expression
found in the canonical ensemble (Chavanis 2002a). The number of nodes in
the profile $\delta\rho$ is determined by the number of intersections
between the spiral in the $(u,v)$ plane and the line $v=2$. We see in
Fig. \ref{uvcrit1} that there is {\it no} intersection for
$\xi\le\alpha=1.78$, where $1.78$ is the stability limit in
GCE. Therefore, collecting the results obtained by different authors,
the perturbation that triggers gravitational instability has two nodes
in MCE (``core-halo'' structure), one node in CE and no node in GCE
(see Fig. \ref{profilpert}).
 
\subsection{The grand microcanonical ensemble}
\label{sec_gmc}

The grand microcanonical ensemble is interesting in its own right but
also because it exhibits a lost-gain-lost stability behaviour similar
to that found for self-gravitating fermions and hard sphere models
(Chavanis 2002c). Now, the stability analysis can be conducted
analytically in GMCE for classical isothermal spheres while only
numerical results are available for fermions and hard sphere
systems. On a technical point of view, GMCE is the most complicated
and richest of all thermodynamical ensembles. We can argue, however,
that GMCE may not be realized in practice. It is not clear, indeed,
what mechanism can be devised so that two systems can share particles
without also sharing energy. One example might be the case of
relativistic particles created and destroyed at equilibrium in a cosmological 
settling.

\begin{figure}
\centering
\includegraphics[width=8.5cm]{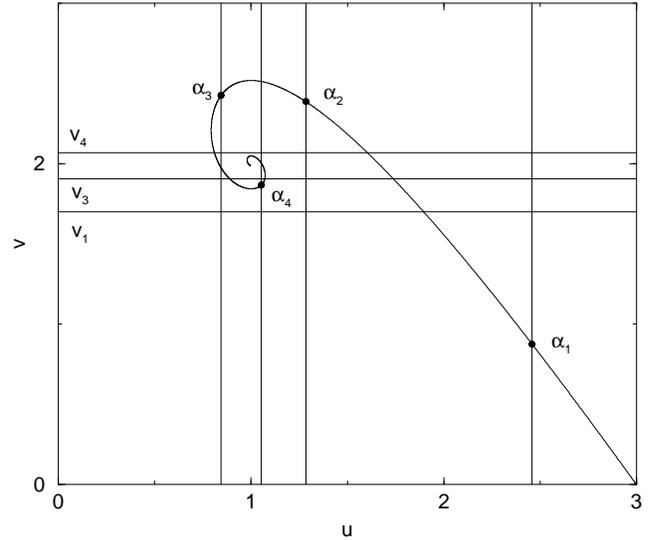}
\caption{Graphical construction determining the number of nodes for the different modes of instability in GMCE. The horizontal lines correspond to the condition (\ref{gmc26}) for the four first critical points.}
\label{nodeuv}
\end{figure}

\begin{figure}
\centering
\includegraphics[width=8.5cm]{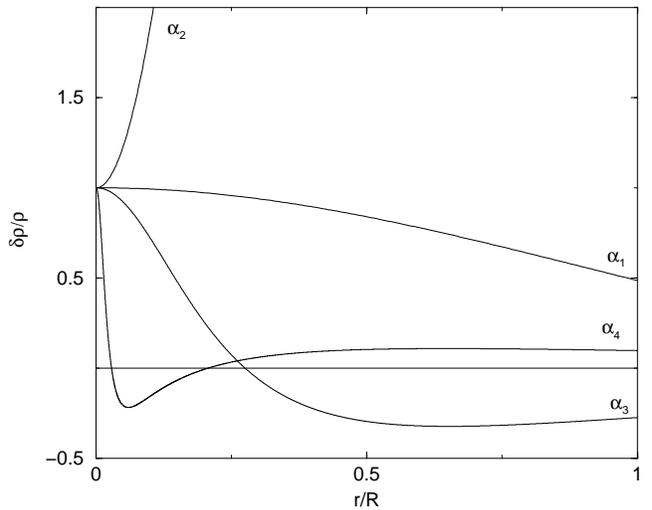}
\caption{Density perturbation profiles for the four first critical points in the series of equilibria in GMCE. We have fixed the constant $c_{1}$ such that $\delta\rho/\rho=1$ at $r=0$. }
\label{prof}
\end{figure}

In GMCE, we have to maximize the thermodynamical potential
\begin{eqnarray}
{{\cal K}\over k}=
\biggl\lbrack {5\over 2}+{3\over 2}\ln\biggl ({2\pi kT\over m}\biggr )+{\mu\over kT}\biggr\rbrack \int{\rho\over m}d^{3}{\bf r}\nonumber\\
-\int{\rho\over m}\ln{\rho\over m}d^{3}{\bf r},\qquad\qquad\qquad\qquad\quad
\label{gmc1}
\end{eqnarray} 
at fixed energy
\begin{equation}
E={3\over 2}kT\int {\rho\over m}d^{3}{\bf r}+{1\over 2}\int\rho\Phi d^{3}{\bf r}.
\label{gmc2}
\end{equation} 
We vary the density profile and use Eq. (\ref{gmc2}) to determine the corresponding variation of temperature. After some algebra, we find that the second order variations of the thermodynamical potential can be written
\begin{eqnarray}
\delta^{2}{\cal K}=-{1\over 2T}\int\delta\rho\delta\Phi d^{3}{\bf r}-{k\over m}\int {(\delta\rho)^{2}\over 2\rho}d^{3}{\bf r}\nonumber\\
-{1\over 3NkT^{2}}\biggl (\int\biggl (\Phi+{3kT\over 2m}\biggr )\delta\rho \ d^{3}{\bf r}\biggr )^{2}.
\label{gmc3}
\end{eqnarray} 
Introducing the variable $X$ defined previously and integrating by parts, we can put the problem in the form
\begin{eqnarray}
\delta^{2}{\cal K}={1\over 2G^{2}}\int_{0}^{R}dr\int_{0}^{R} dr' X'(r)K(r,r')X'(r'),
\label{gmc5}
\end{eqnarray} 
with
\begin{eqnarray}
K(r,r')=\biggl\lbrace {G\over T}+{k\over m}{d\over dr}\biggl ({1\over 4\pi \rho}{d\over dr}\biggr )\biggr\rbrace\delta(r-r')\nonumber\\
-{2\over 3NkT^{2}}\biggl \lbrack \biggl (\Phi+{3kT\over 2m}\biggr ) r\biggr\rbrack'(r)\biggl \lbrack \biggl (\Phi+{3kT\over 2m}\biggr ) r\biggr\rbrack'(r').
\label{gmc6}
\end{eqnarray} 
The stability analysis reduces therefore to the study of the eigenvalue equation
\begin{eqnarray}
\int_{0}^{R} dr' K(r,r')X_{\lambda}'(r')=\lambda X_{\lambda}'(r).
\label{gmc7}
\end{eqnarray} 
The condition of marginal stability ($\lambda=0$) reads
\begin{eqnarray}
{GX'\over T}+{k\over m}{d\over dr}\biggl ({1\over 4\pi \rho}{dX'\over dr}\biggr )=
{2\over 3NkT^{2}}\biggl \lbrack \biggl (\Phi+{3kT\over 2m}\biggr ) r\biggr\rbrack' \nonumber\\
\times\int_{0}^{R} \biggl \lbrack \biggl (\Phi+{3kT\over 2m}\biggr ) r\biggr\rbrack'   X'dr.\qquad\qquad\qquad
\label{gmc8}
\end{eqnarray} 
Integrating once, we obtain
\begin{eqnarray}
{GX\over T}+{k\over m}{X''\over 4\pi\rho}=
{2\over 3NkT^{2}} \biggl (\Phi+{3kT\over 2m}\biggr ) r \nonumber\\
\times\int_{0}^{R} \biggl \lbrack \biggl (\Phi+{3kT\over 2m}\biggr ) r\biggr\rbrack'   X'dr.\qquad
\label{gmc9}
\end{eqnarray} 
The gravitational potential $\Phi$ is related to the Emden function $\psi(\xi)$ by the relation $\psi=\beta(\Phi-\Phi_{0})$. Using $\Phi(R)=-GM/R$ and Eq. (\ref{eq2}), we get
\begin{equation}
\beta\Phi=\psi-\psi(\alpha)-v_{0}.
\label{gmc10}
\end{equation} 
Inserting this relation in Eq. (\ref{gmc9}) and introducing the dimensionless variables defined in Sec. \ref{sec_emden}, we find that
\begin{eqnarray}
X+e^{\psi}X''={1\over\alpha v_{0}}\biggl\lbrack \psi-\psi(\alpha)-v_{0}+{3\over 2}\biggr\rbrack\xi\nonumber\\
\times\biggl\lbrace -{2\over 3}\int_{0}^{\alpha}X\xi e^{-\psi}d\xi+X(\alpha)\biggr\rbrace,
\label{gmc11}
\end{eqnarray} 
where we have integrated by parts and used the identity $(\xi\psi)''=\xi e^{-\psi}$ equivalent to the Emden equation (\ref{emden2}). Writing 
\begin{equation}
V={2\over 3v_{0}\alpha}\biggl ({3\over 2}X(\alpha)-\int_{0}^{\alpha}X\xi e^{-\psi}d\xi\biggr ),
\label{gmc12}
\end{equation} 
\begin{equation}
W=-\biggl\lbrack \psi(\alpha)+v_{0}-{3\over 2}\biggr\rbrack V,
\label{gmc13}
\end{equation} 
we are led to solve the problem
\begin{equation}
e^{\psi}X''+X=V\psi\xi+W\xi,
\label{gmc14}
\end{equation} 
\begin{equation}
X(0)=X'(\alpha)=0.
\label{gmc15}
\end{equation} 
Seeking a solution of Eq. (\ref{gmc14}) in the form
\begin{equation}
X=a_{1}\xi+a_{2}\xi^{2}\psi'+a_{3}\xi\psi,
\label{gmc16}
\end{equation} 
and using the identities  (\ref{lb2}), we find that
\begin{equation}
X=(W-V)\xi+b(\xi^{2}\psi'-2\xi)+V\xi\psi.
\label{gmc18}
\end{equation}
The constant $b$ is determined by the boundary condition $X'(\alpha)=0$. We thus find that the general solution of Eqs. (\ref{gmc14}) and (\ref{gmc15}) is
\begin{eqnarray}
X(\xi)=-\biggl\lbrack \psi(\alpha)+v_{0}-{1\over 2}\biggr\rbrack V\xi+V\xi\psi\nonumber\\
-{V\over 2}{1\over u_{0}v_{0}-2}(\xi^{2}\psi'-2\xi), 
\label{gmc19}
\end{eqnarray}
where we have used Eq. (\ref{gmc13}).  The values of $\alpha$ at which
a change of stability occurs ($\lambda=0$) are obtained by inserting
the solution (\ref{gmc19}) for $X$ in Eq. (\ref{gmc12}). We need the
identities
\begin{equation}
\int_{0}^{\alpha}\xi^{2}e^{-\psi}d\xi=\alpha v_{0},
\label{gmc20}
\end{equation}
\begin{equation}
\int_{0}^{\alpha}\xi^{3}\psi'e^{-\psi}d\xi=\alpha v_{0}(3-u_{0}),
\label{gmc21}
\end{equation}
\begin{equation}
\int_{0}^{\alpha}\xi^{2}\psi e^{-\psi}d\xi=\alpha v_{0}(\psi(\alpha)+v_{0}-6+2 u_{0}).
\label{gmc22}
\end{equation}
which are derived in Appendix \ref{sec_id1}.  After simplification, we
obtain the condition
\begin{equation}
8u_{0}^{2}v_{0}-10u_{0}v_{0}-17u_{0}+21=0,
\label{gmc23}
\end{equation}
which coincides with the condition (\ref{eq9}) deduced from the
turning point criterion. The perturbation profile at the critical
points at which $\lambda=0$ is given by Eq. (\ref{q1}) with
Eq. (\ref{gmc19}) for $X$. Using Eqs. (\ref{emden1}), (\ref{gmc14})
and (\ref{gmc19}), we get
\begin{equation}
{\delta\rho\over\rho}={c_{1}\over 4\pi}(2u_{0}v_{0}-6+v),
\label{gmc25}
\end{equation} 
where $c_{1}$ is a constant. The nodes of the perturbation profile are determined by the condition
\begin{equation}
v(\xi_{i})=6-2u_{0}v_{0},
\label{gmc26}
\end{equation} 
with $\xi_{i}\le\alpha$.  The number of nodes can be found by a graphical
construction. In Fig. \ref{nodeuv}, we plot the lines defined by
Eq. (\ref{gmc26}) in the $(u,v)$ plane for the four first critical
points in the series of equilibria. The corresponding perturbation
density profiles are plotted in Fig. \ref{prof}. The first mode of
instability $\alpha_{1}$ has no node. The profile of the second mode
$\alpha_{2}$ {\it increases} (in absolute value) with distance. The
absence of node at $\alpha_{2}$ is a signature of the fact that
stability is regained at that point. The mode $\alpha_{3}$ at which
stability is lost again has one node, the mode $\alpha_{4}$ two nodes
etc...

\section{Connexion with statistical mechanics}
\label{sec_field}

In this section, we briefly discuss the relation between the
thermodynamics and the statistical mechanics (and field theory) of
self-gravitating systems. We shall just discuss the physical ideas
without entering into technical details.

In the microcanonical ensemble, the object of fundamental interest is
the density of states
\begin{eqnarray}
g(E)=\int \delta (E-H)\prod_{i=1}^{N}d^{3}{\bf r}_{i}d^{3}{\bf v}_{i},
\label{field1}
\end{eqnarray} 
where 
\begin{equation}
H={1\over 2}\sum_{i=1}^{N}mv_{i}^{2}-\sum_{i\le j}{Gm^{2}\over |{\bf r}_{i}-{\bf r}_{j}|},
\label{field2}
\end{equation} 
is the Hamiltonian of the self-gravitating gas. By definition, $g(E)dE$ is proportional to the number of {\it microstates} (specified by the position and the velocity of all $N$ particles) with energy between $E$ and $E+dE$. The entropy
is defined by
\begin{equation}
S(E)=\ln g(E).
\label{field2bis}
\end{equation} 
As is customary in statistical mechanics, we decompose phase space
into macrocells and divide these macrocells into a large number of
microcells. A {\it macrostate} is characterized by the number of
particles in each macrocell (irrespective of their precise position
and velocity in the cell), or equivalently by the smooth distribution
function $f({\bf r},{\bf v})$. If we call $W_{i}$ the number of
microstates corresponding to the macrostate $i$, we can rewrite
Eq. (\ref{field1}) formally in the form
\begin{eqnarray}
g(E)=\sum_{E_{i}=E}W_{i}=\sum_{E_{i}=E}e^{S_{i}}=\sum_{E_{i}=E}e^{Ns_{i}},
\label{field3}
\end{eqnarray} 
where the sum runs over all macrostates with energy $E$ (and mass
$M$). In addition, $S_{i}=\ln W_{i}$ is the Boltzmann entropy
(\ref{ens4}) of the macrostate $i$ (i.e., $S_{i}\equiv S[f]$) and
$s_{i}=S_{i}/N$ the entropy per particle. The thermodynamic limit for
self-gravitating systems corresponds to $N\rightarrow +\infty$ in such
a way that all the dimensionless control parameters defined in
Sec. \ref{sec_eq} remain finite. This can be written for example
$N,R\rightarrow +\infty$ with $N/R$ finite (and $E/N\sim 1$,
$\beta\sim 1$). This is a very unusual thermodynamic limit due to the
long-range nature of the interactions.
 Now, at the thermodynamic limit, the sum in Eq. (\ref{field3})
is dominated by the macrostate $f_{*}$ which maximizes the Boltzmann
entropy (\ref{ens4}) at fixed mass and energy:
\begin{eqnarray}
g(E)\simeq e^{S[f_{*}]},\qquad S(E)=S[f_{*}].
\label{field4}
\end{eqnarray}
It has to be noted that the sum in Eq. (\ref{field3}) is sharply
peaked around $f_{*}$, which corresponds to what is called a large
deviation property in mathematics. Therefore, the mean-field approximation
is {\it exact} for self-gravitating systems at the thermodynamic limit. 

In fact, if we look at the problem more carefully, the argument
leading to Eq. (\ref{field4}) is not so obvious because the entropy
$S[f]$ has no global maximum, except if we introduce small-scale and
large-scale cut-offs. The introduction of these cut-offs is necessary
to define correctly the density of state (\ref{field1}), otherwise the
integral would diverge (Padmanabhan 1990). The divergence at large
scales is associated to the fact that a stellar system has the
tendency to evaporate under the effect of encounters.  In
reality, stellar systems are not allowed to extend to infinity because
they interact with the surrounding.  For example, globular clusters
are subject to the tides of a nearby galaxy so that the Boltzmann
distribution has to be truncated at high energies. The Michie-King
model,
\begin{eqnarray}
{f}=\Biggl\lbrace \begin{array}{cc}
A {(e^{-\beta\epsilon}-e^{-\beta\epsilon_{m}})}   & \epsilon< \epsilon_{m}, \\
0  & \epsilon\ge \epsilon_{m},
\end{array}
\label{mk}
\end{eqnarray}  
which is a truncated isothermal, can take into account the evaporation
of high energy stars and provides a good description of about $80\%$
of globular clusters (Binney \& Tremaine 1987). It can be derived from
the Fokker-Planck equation (taking into account the encounters between
stars) by imposing that the distribution function vanishes at the
escape energy $\epsilon=\epsilon_{m}$. In that case, the system is not
truly static since it gradually loses stars but we can consider that a
globular cluster passes by a succession of quasi-equilibrium
configurations.  Instead of working with truncated models, it is more
convenient to confine the system within a box of radius $R$, the box
radius playing the role of a tidal radius. The divergence at small
scales of the density of states reflects the natural tendency of a
stellar system to form binaries. In this extreme case, the size $a$ of
the stars provides a small-scale cut-off.  The influence of a
small-scale cut-off on the form of the equilibrium phase diagram is
discussed by Chavanis (2002c) and Chavanis \& Ispolatov (2002). When
the small-scale cut-off $a\rightarrow 0$, which is the limit relevant
for globular clusters since the size of the stars is in general much
smaller than the interstellar distance, the entropy $S[f]$ possesses
two maxima (for sufficiently high energies). The local entropy maximum
describes a ``gaseous'' phase insensitive to the small scale
cut-off. It corresponds to the isothermal spheres studied
previously. The global entropy maximum is dominated by the small scale
cut-off and corresponds to a ``condensed'' phase made of a few stars
close together (or simply a single binary) surrounded by a diffuse
halo. This ``core-halo'' structure is the most probable configuration
of a stellar system so it is expected to be reached at equilibrium.
Since the observed globular clusters, described by the Michie-King
model, do not possess this core-halo structure, we conclude that they
are not in true equilibrium but that they are slowly evolving in that
direction. However, the formation of binaries can take extremely long
times, much larger than the age of the universe. Indeed, equilibrium
statistical mechanics tells nothing about the relaxation time.  For
the timescales contemplated in astrophysics, globular clusters can be
considered as {\it metastable} equilibrium states corresponding to
local entropy maxima. Indeed, the probability of transition from a
gaseous state to a condensed state (corresponding to the formation of
a binary) is extremely small (Katz \& Okamoto 2000, Chavanis
\& Ispolatov 2002).  Therefore, in the calculation of the density of
states (\ref{field1}), we must discard {\it by hands} the states which
correspond to collapsed configurations (binaries) since they cannot be
reached in the timescales of interest. Accordingly, Eq. (\ref{field4})
is correct with $f_{*}$ being the {\it local} maximum of $S[f]$. This is
precisely the situation studied in Secs. \ref{sec_thermo} and
\ref{sec_stab}. In fact, this picture is correct only for globular
clusters with high energy $E$, well above the Antonov limit
$E_{c}$. Because of evaporation, the energy of a globular cluster
slowly decreases until it reaches the point at which the system cannot
be in equilibrium anymore (e.g., Katz 1980). At that point, an isothermal
sphere becomes unstable and the system undergoes a gravitational
collapse (gravothermal catastrophe). The evolution proceeds
self-similarly with the formation of a high density core with a
shrinking radius and a small mass (H\'enon 1961, Cohn 1980, Lynden-Bell \&
Eggleton 1980). This core collapse concerns typically $20\%$ of
globular clusters. In theory, the central density becomes infinite in
a finite time. In practice, the formation of binaries can release
sufficient energy to stop the collapse (H\'enon 1961) and even drive a
reexpansion of the system (Inagaki \& Lynden-Bell 1983). Then, a
series of gravothermal oscillations should follow (Bettwieser \&
Sugimoto 1984).

In the canonical ensemble, the object of interest is the partition function
\begin{eqnarray}
Z(\beta)=\int e^{-\beta H}\prod_{i=1}^{N}d^{3}{\bf r}_{i}d^{3}{\bf v}_{i}.
\label{field5}
\end{eqnarray}  
The free energy is defined by
\begin{eqnarray}
F(\beta)=-{1\over\beta}\ln Z.
\label{field6}
\end{eqnarray} 
The partition function is related to the density of states by the Laplace transform 
\begin{eqnarray}
Z(\beta)=\int_{-\infty}^{+\infty}dE\ g(E)e^{-\beta E}.
\label{field7}
\end{eqnarray} 
Using Eq. (\ref{field3}), it can be rewritten
\begin{eqnarray}
Z(\beta)=\sum_{i}e^{S_{i}-\beta E_{i}}=\sum_{i}e^{J_{i}},
\label{field8}
\end{eqnarray}
where $J_{i}=S_{i}-\beta E_{i}$ is the Massieu function and the sum runs over all macrostates with mass $M$. In the thermodynamic limit, the partition function is dominated by the contribution of the macrostate which maximizes the Massieu function $J_{i}\equiv J[f]$ at fixed mass. In this limit, the mean field approximation is exact and we have
\begin{eqnarray}
Z(\beta)\simeq e^{J[f_{*}]},\qquad F=E-TS.
\label{field9}
\end{eqnarray}
Again, the Massieu function $J[f]$ has no global maximum in the
absence of cut-off. If we introduce a small scale cut-off $a$ and a
large scale cut-off $R$ and consider the limit $a/R\ll 1$, it is found
that the Massieu function $J[f]$ has two maxima (for sufficiently high
temperatures). The local maximum corresponds to isothermal ``gaseous''
configurations and the global maximum to a ``condensed'' structure in
which all the particles are packed together (Aronson \& Hansen 1972,
Chavanis 2002c, Chavanis \& Ispolatov 2002). As $a\rightarrow 0$, the
density profile of the condensed state tends to a Dirac peak
(Kiessling 1989). As in the microcanonical ensemble, the ``gaseous''
configurations are metastable but they can be long lived and relevant
for astrophysical purposes. This justifies why we only select the
contribution of the local maximum of $J[f]$ in
Eq. (\ref{field9}). This treatment is, however, valid only for
sufficiently high temperatures. Below $T_{c}$, a phase transition
similar to the gravothermal catastrophe occurs in the canonical
ensemble and leads to a condensed structure containing almost all the
mass.  The inequivalence of statistical ensembles regarding the
formation of binaries (in MCE) or Dirac peak (in CE) is further
discussed in Appendices A and B of Sire \& Chavanis (2002).

Finally, the grand canonical partition function is defined by
\begin{eqnarray}
Z_{GC}=\sum_{N=0}^{+\infty}{z^{N}\over N!}\int\prod_{i=1}^{N}d^{3}{\bf r}_{i}d^{3}{\bf v}_{i}e^{-\beta H_{N}},
\label{field10}
\end{eqnarray}
where $z$ is the fugacity. Applying the standard Hubbard-Stratanovich
transformation, it can be rewritten in the form of a path integral
(Horwitz \& Katz 1978, Padmanabhan 1991, de Vega \& Sanchez 2002)
\begin{eqnarray}
Z_{GC}=\int{\cal D}\phi\ e^{-{1\over kT_{eff}}\int d^{3}{\bf r}\lbrace {1\over 2}(\nabla\phi)^{2}-\mu^{2}e^{\phi}\rbrace},
\label{field11}
\end{eqnarray}
\begin{eqnarray}
T_{eff}=4\pi {Gm^{2}\over T},\qquad \mu^{2}=\sqrt{2\over \pi}zGm^{7/2}\sqrt{T}.
\label{field12}
\end{eqnarray}
for the Liouville action
\begin{eqnarray}
A\lbrack\phi\rbrack=-{1\over kT_{eff}}\int d^{3}{\bf r}\lbrace {1\over 2}(\nabla\phi)^{2}-\mu^{2}e^{\phi}\rbrace.
\label{field13}
\end{eqnarray}
It should be emphasized that $\phi$ is a {formal} field which is {\it
not} the gravitational field. In the mean-field approximation, the path
integral is dominated by the contribution of the field $\phi_{0}$
which maximizes the Liouville action (\ref{field13}). The cancellation
of the first order variations gives
\begin{eqnarray}
\Delta\phi_{0}=-\mu^{2}e^{\phi_{0}},
\label{field14}
\end{eqnarray}
which is similar to the Boltzmann-Poisson equation
(\ref{emden0}). This allows one to identify $\phi_{0}$ with the
equilibrium gravitational potential  $\Phi$ (up to a negative proportionality
factor). The second order variations of $A$ at the critical point must
be negative for $A[\phi_{0}]$ to be a maximum. This yields the
condition
\begin{eqnarray}
\int d^{3}{\bf r}\ \delta\phi\lbrace \Delta+\mu^{2}e^{\phi_{0}}\rbrace\delta\phi\le 0,
\label{field15}
\end{eqnarray}
which is equivalent to the maximization of the grand potential $G$
(see Sec. \ref{sec_gc}). Therefore, the thermodynamical approach is
equivalent to the statistical mechanics or field theory approach. We
note, however, that the Liouville action $A[\phi]$ is not the same
functional as the grand potential $G[\rho]$ although they give the
same critical points and the same conditions of stability.

\section{Generalized thermodynamics and Tsallis entropy}
\label{sec_tsallis}

\subsection{General considerations}
\label{sec_cons}

It has been recently argued that the classical Boltzmann entropy may not be relevant for non-extensive systems and that Tsallis entropies, also called $q$-entropies, should be used instead.   These entropies can be written
\begin{equation}
S_{q}=-{1\over q-1}\int ({f}^{q}-{f}) d^{3}{\bf r}d^{3}{\bf v},
\label{sp1}
\end{equation} 
where  $q$ is a real number. For $q\rightarrow 1$,
this expression returns the classical Boltzmann entropy (\ref{ens4})
as a particular case. 

The first application of Tsallis generalized thermodynamics in stellar
systems was made by Plastino \& Plastino (1997) who showed that the
extremization of Tsallis entropy leads to stellar polytropes with an index
\begin{equation}
n={3\over 2}+{1\over q-1}.
\label{cons}
\end{equation} 
For $q\rightarrow 1$, or $n\rightarrow +\infty$, we recover isothermal
spheres. For $q>9/7$ (i.e. $n<5$), the density drops to zero at a finite distance so
that the mass is finite. Then, Taruya \& Sakagami (2002a) considered
the stability problem in the microcanonical ensemble by extending
Padmanabhan's classical analysis of the Antonov instability to the
case of polytropic distribution functions. A configuration is stable
in the sense of generalized thermodynamics if it corresponds to a
maximum of Tsallis entropy at fixed mass and energy. They considered
polytropic spheres confined within a box and showed that polytropes
with index $n\ge 5$ become unstable above a certain density contrast
while polytropes with $n\le 5$ are always stable. On the other hand,
if we consider the canonical ensemble, we have to select maxima of
Tsallis {free energy} $J_{q}=S_{q}-\beta E$ at fixed mass and $\beta$
(the generalized inverse temperature).  It is found (Taruya
\& Sakagami 2002b; see also Sec. \ref{sec_TsallisC}) that polytropes with index $n\ge 3$ become unstable above a certain density contrast while
polytropes with $n\le 3$ are always stable.  In our sense, the main
interest of the $q$-entropies (and their attractive nature) is to
offer a simple generalization of classical thermodynamics which leads
to models that are still {analytically tractable} since exponentials
are replaced by power laws. This approach is interesting to develop, at
least formally, as it offers a nice connexion between polytropic and
isothermal spheres, which are the most popular mathematical models of
self-gravitating systems. We shall therefore devote a section to this
generalization. However, in Sec. \ref{sec_st}, we shall argue that
Tsallis entropies are just a particular case of $H$-functions $S[f]$
whose maximization at fixed mass and energy determines nonlinearly
stable stationary solutions of the Vlasov equation (Tremaine et al
1986). We shall also argue that the maximization of the $J$-function
$J[f]=S-\beta E$ at fixed mass determines nonlinearly stable
stationary solutions of the Euler-Jeans equations. This discussion
should demystify the concept of generalized thermodynamics introduced
by Tsallis (1988).

\subsection{Stellar polytropes}
\label{sec_sp}  

The extremization of Tsallis entropy at fixed mass and energy yields (Plastino \& Plastino 1997,Taruya \& Sakagami 2002a)
\begin{equation}
{f}=A\biggl\lbrack \alpha-\Phi-{v^{2}\over 2}\biggr \rbrack^{1\over q-1},
\label{sp2}
\end{equation} 
\begin{equation}
A=\biggl\lbrack {(q-1)\beta\over q}\biggr \rbrack^{1\over q-1}, \qquad \alpha={1+(q-1)(\mu/T)\over (q-1)\beta},
\label{sp3}
\end{equation} 
where $\beta=1/T$ and $\mu/T$ (related to $\alpha$) are Lagrange
multipliers which can be called inverse temperature and chemical
potential in the generalized sense. The distribution function
(\ref{sp2}) corresponds to stellar polytropes that were first
introduced by Plummer (1911). Their derivation from a variational
principle was discussed by Ipser (1974) among others. Note that
Eq. (\ref{sp2}) is valid for $v^{2}\le 2(\alpha-\Phi)$. For $v^{2}\ge
2(\alpha-\Phi)$, we set $f=0$. The spatial density ${\rho}=\int f
d^{3}{\bf v}$ and the pressure $p={1\over 3}\int fv^{2}d^{3}{\bf v}$
can be expressed as
\begin{equation}
{\rho}=4\sqrt{2}\pi A(\alpha-\Phi)^{n}B ({3/2},n-{1/2}  ),
\label{sp4}
\end{equation} 
\begin{equation}
{p}={4\sqrt{2}}\pi A (\alpha-\Phi)^{n+1}{1\over n+1}  B
({3/2},n-{1/2} ),
\label{sp5}
\end{equation} 
with $B(a,b)$ being the $\beta$ function and use has been made of
Eq. (\ref{cons}). Furthermore, in obtaining Eq. (\ref{sp5}), we have
used the identity $B(m+1,n)=mB(m,n)/(m+n)$. The integrability
condition $\int f d^{3}{\bf v}<\infty$ requires that
$n>1/2$. Eliminating the gravitational potential between the relations
(\ref{sp4}) and (\ref{sp5}), we recover the well-known fact that
stellar polytropes satisfy the equation of state
\begin{equation}
p=K\rho^{\gamma},\qquad \gamma=1+{1\over n},
\label{sp6}
\end{equation} 
like gaseous polytropes (note, however, that these systems do not have
the same phase space distribution function, unlike isothermal
spheres). In the present context, the polytropic constant is given by
\begin{equation}
K={1\over (n+1)}\biggl\lbrace 4\sqrt{2}\pi A B ({3/2},n-{1/2})\biggr\rbrace^{-1/n}.
\label{sp7}
\end{equation}

Using Eqs. (\ref{sp2}) and (\ref{sp4}), the distribution function can be written as a function of the density as
\begin{eqnarray}
{f}={1\over Z}\biggl \lbrack \rho^{1/n}-{v^{2}/2\over (n+1)K}\biggr \rbrack^{n-3/2},
\label{sp8}
\end{eqnarray} 
with
\begin{eqnarray}
Z={4\sqrt{2}\pi B(3/2,n-1/2)}{\lbrack
K(n+1)\rbrack^{3/2}}.
\label{sp8bis}
\end{eqnarray}

Using the foregoing relations, it is possible to express the total energy and the entropy in terms of $\rho$ and $p$ according to (Taruya \& Sakagami 2002b; see also Sec. \ref{sec_dsgs})
\begin{equation}
E={3\over 2}\int pd^{3}{\bf r}+{1\over 2}\int\rho\Phi d^{3}{\bf r},
\label{sp9}
\end{equation} 
\begin{equation}
S=-\biggl (n-{3\over 2}\biggr )\biggl\lbrace {1\over T}\int p d^{3}{\bf r}-{M}\biggr\rbrace.
\label{sp10}
\end{equation}

\subsection{The Lane-Emden equation}
\label{sec_lane}

The equilibrium structure of a polytrope is determined by substituting the relation (\ref{sp4}) between  $\rho$  and the gravitational potential $\Phi$ inside the Poisson equation (\ref{ens3}). This is equivalent to using  the condition of hydrostatic equilibrium $\nabla p=-\rho\nabla\Phi$ with the equation of state (\ref{sp6}). Letting
\begin{equation}
\rho=\rho_{0}\theta^{n},\qquad \xi=\biggl\lbrack {4\pi G\rho_{0}^{1-1/n}\over K(n+1)}\biggr\rbrack^{1/2}r,
\label{lane1}
\end{equation} 
where $\rho_{0}$ is the central density, we can reduce the condition of hydrostatic equilibrium to the Lane-Emden equation (Chandrasekhar 1942)
\begin{equation}
{1\over\xi^{2}}{d\over d\xi}\biggl (\xi^{2}{d\theta\over d\xi}\biggr )=-\theta^{n},
\label{lane2}
\end{equation} 
with boundary conditions $\theta(0)=1$, $\theta'(0)=0$. 

For $n<5$, the density vanishes at some radius identified with the
radius of the polytrope (in terms of the normalized distance, this
corresponds to $\theta=0$ at $\xi=\xi_{1}$). Such polytropic
configurations (with infinite density contrasts) will be refered to as
{\it complete} polytropes. We shall also consider the case of {\it
incomplete} polytropes, with arbitrary index,  enclosed within a box of radius
$R$. For these configurations, the Lane-Emden equation must be solved
between $\xi=0$ and $\xi=\alpha$ with
\begin{equation}
\alpha=\biggl\lbrack {4\pi G\rho_{0}^{1-1/n}\over K(n+1)}\biggr\rbrack^{1/2}R.
\label{lane3}
\end{equation} 

We also recall the expression of the Milne variables that will be used in the sequel, 
\begin{equation}
u=-{\xi\theta^{n}\over \theta'},\qquad v=-{\xi\theta'\over\theta}.
\label{lane4}
\end{equation} 
These variables satisfy the identities
\begin{equation}
{1\over u}{du\over d\xi}={1\over\xi}(3-nv-u),
\label{lane5}
\end{equation} 
\begin{equation}
{1\over v}{dv\over d\xi}={1\over \xi}(u+v-1),
\label{lane6}
\end{equation} 
which can be derived from Eq. (\ref{lane2}). The phase portrait of
polytropes in the $(u,v)$ plane and the properties that we shall need in
the following are summarized in Chavanis (2002b).

\subsection{Turning point criterion}
\label{sec_tpc}  

Eliminating $A$ between Eqs. (\ref{sp3}) and (\ref{sp7}), we find that 
\begin{equation}
\beta=\biggl\lbrack {4\pi Q^{n-1}\over K^{n}}\biggr\rbrack^{2\over 2n-3},
\label{tpc1}
\end{equation} 
where we have defined (Taruya \& Sakagami 2002b)
\begin{equation}
\qquad Q=\biggl\lbrace {\bigl (n-{1\over 2}\bigr )^{n-{3\over 2}}\over 16\sqrt{2}\pi^{2}B(3/2,n-1/2)(n+1)^{n}}\biggr\rbrace^{1\over n-1}.
\label{tpc2}
\end{equation} 
In a preceding paper (Chavanis 2002b), we have established that the mass of the configuration is related to the central density $\rho_{0}$ (through the parameter $\alpha$) by the relation
\begin{equation}
\eta\equiv {M\over 4\pi}\biggl\lbrack {4\pi G\over K(1+n)}\biggr\rbrack^{n\over n-1}{1\over R^{n-3\over n-1}}=(u_{0}v_{0}^{n})^{1\over n-1}.
\label{tpc3}
\end{equation} 
We had conjectured that $\eta$ plays the role of a temperature in the context of Tsallis generalized thermodynamics. This is indeed the case since, using  Eq. (\ref{tpc1}), $\eta$ can be expressed in terms of $\beta$ as 
\begin{equation}
\eta={1\over Q}\biggl\lbrace {G^{n}M^{n-1}\over R^{n-3}}\beta^{n-{3\over 2}}{1\over (n+1)^{n}}\biggr\rbrace^{1\over n-1}.
\label{tpc4}
\end{equation} 
For $n\rightarrow +\infty$, this parameter becomes equivalent to the parameter $\eta=\beta GM/R$  introduced in the study of isothermal spheres (Chavanis 2002a). 

In the framework of generalized thermodynamics, the parameter conjugate to the energy with respect to Tsallis entropy $S_{q}$ is the inverse temperature $\beta$ (microcanonical description) and the parameter conjugate to the inverse temperature with respect to Tsallis free energy $J_{q}=S_{q}-\beta E$ is $-E$ (canonical description). The $E-T$ curve has been plotted and discussed in Chavanis (2002b) for different values of the polytropic index. According to the turning point criterion, the series of equilibria becomes unstable in the canonical ensemble for $d\eta/d\alpha=0$. We have shown  that this condition is equivalent to
\begin{equation}
u_{0}={n-3\over n-1}=u_{s}.
\label{tpc5}
\end{equation} 
For $n\le 3$, the polytropes are always stable since the
$\eta(\alpha)$ curve is monotonic. The polytropes start to be unstable
(for sufficiently high density contrasts) when $n>3$.  This
generalized thermodynamical stability criterion coincides with the
Jeans dynamical stability criterion based on the Navier-Stokes
equations for a polytropic gas (Chavanis 2002b).  Since $K$ is
considered as a constant when we analyze the stability of polytropic
spheres with respect to the Navier-Stokes equations, and since $K$ is
related to the temperature via Eq. (\ref{tpc1}), this corresponds to a
sort of ``canonical situation''. This explains why the condition of
dynamical stability for gaseous polytropes is equivalent to the
condition of thermodynamical stability for stellar polytropes in the
canonical ensemble. This equivalence is further discussed in
Sec. \ref{sec_st}.

\begin{figure}
\centering
\includegraphics[width=8.5cm]{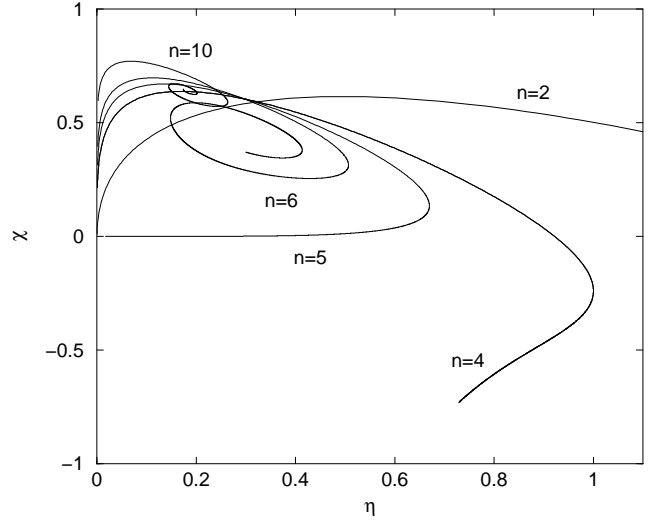}
\caption{Fugacity vs mass (for a given temperature) in the framework of generalized thermodynamics for different polytropic index. For $1<n\le 3$, $\chi$ has one turning point but not $\eta$; for $3<n\le 5$, both $\chi$ and $\eta$ have one turning point; for $n>5$, $\chi$ and $\eta$ have an infinity of turning points. For $n<5$, complete polytropes correspond to the terminal point of the $\eta-\chi$ curve. For $n>5$, the curve has a spiral behaviour towards the limit point $({\eta}_{s},{\chi}_{s})$ corresponding to the singular sphere $(u_{s},v_{s})=({n-3\over n-1},{2\over n-1})$.}
\label{etachipoly}
\end{figure}

We now turn to the grand canonical situation. Combining
Eqs. (\ref{sp4}), (\ref{sp3}) and (\ref{tpc2}), we can write the density
field as
\begin{equation}
\rho={Q^{1-n}\over 4\pi (n+1)^{n}\beta^{3/2}}\biggl\lbrack n-{3\over 2}+{\mu\over T}-\beta\Phi\biggr\rbrack^{n}.
\label{tpc5bis}
\end{equation}
This expression suggests to define a generalized fugacity  by the relation $n\ z^{1/n}\equiv n-3/2+\mu/T$ so that $z$ is equivalent to the usual fugacity $e^{\mu\over T}$ when $n\rightarrow +\infty$. Taking $r=R$ in Eq. (\ref{tpc5bis}) and comparing with Eq. (\ref{lane1}), we get 
\begin{equation}
\rho_{0}\theta(\alpha)^{n}={Q^{1-n}\over 4\pi (n+1)^{n}\beta^{3/2}}  \biggl (n\ z^{1/n}+{\beta GM\over R}\biggr )^{n},
\label{tpc6}
\end{equation}
where we have used $\Phi(R)=-GM/R$. Expressing the central density in terms of $\alpha$, using Eqs. (\ref{lane3}) and (\ref{tpc1}), we find
\begin{equation}
\rho_{0}={Q\over 4\pi}\biggl \lbrack {(n+1)^{n}\alpha^{2n}\over G^{n}R^{2n}\beta^{n-{3\over 2}}}\biggr\rbrack^{1\over n-1}.
\label{tpc7}
\end{equation}
From Eqs. (\ref{tpc6}) and (\ref{tpc7}), we get
\begin{equation}
n\ z^{1/n}=Q\biggl\lbrack {(n+1)^{n}\beta^{1/2}\over GR^{2}}\biggr\rbrack^{1\over n-1}\alpha^{2\over n-1}\theta(\alpha)-{\beta GM\over R}.
\label{tpc8}
\end{equation}
Eliminating $M$ thanks to Eq. (\ref{tpc4}) and introducing the Milne variables (\ref{lane4}), we finally obtain
\begin{equation}
\chi\equiv {n\over Q}z^{1/n}\biggl\lbrack {GR^{2}\over (n+1)^{n}\beta^{1/2}}\biggr\rbrack^{1\over n-1}=(u_{0}v_{0})^{1\over n-1}(1-v_{0}).
\label{tpc10}
\end{equation}
For $n\rightarrow +\infty$, this parameter becomes equivalent to the
one introduced in Sec. \ref{sec_eq} for isothermal spheres (note that
for sufficiently small index $n$, the parameter $\chi$ can become
negative when $v_{0}>1$; therefore, the definition of $z$ given
previously is not valid for small $n$). According to the turning
point criterion, the series of equilibria becomes unstable in the
grand canonical ensemble when $d\chi/d\alpha=0$. This gives the
condition
\begin{equation}
2(1-v_{0})-(n-1)u_{0}v_{0}=0.
\label{tpc11}
\end{equation}
The values of $\alpha$ at which $\chi$ is extremum can be determined
by a graphical construction like in Chavanis (2002b). They correspond to the
 intersections between the hyperbole defined by Eq. (\ref{tpc11})
and the solution curve in the $(u,v)$ plane (which depends on the index $n$). From these graphical
constructions, we deduce that the function $\chi(\alpha)$ has one maximum for
$n\le 5$. For $n=5$, the Lane-Emden equation can be solved
analytically and we get
\begin{equation}
\chi={\alpha^{4}\over (1+{1\over 3}\alpha^{2})^{12}},
\label{tpc12}
\end{equation}
which is maximum for $\alpha=(3/5)^{1/2}$. For $n>5$, the function
$\chi(\alpha)$ has an infinite number of extrema. The onset of
gravitational instability corresponds to the first maximum. The
fugacity vs mass plot (for a given temperature) can be deduced from
Fig. \ref{etachipoly}. Collecting the results obtained by different
authors, incomplete polytropes with high density contrasts are
unstable in MCE for $n>5$, in CE for $n>3$ and in GCE for
all $n$. Complete polytropes are always stable in MCE, stable in CE only
for $n<3$ and never stable in GCE.

In the following, we analyze the stability of polytropic spheres by
explicitly calculating the second order variations of the
thermodynamical potential. This study has been done by Taruya
\& Sakagami (2002a,b) in MCE and CE. We shall consider CE in relation
with our previous study (Chavanis 2002b), GCE and GMCE.

\subsection{Generalized thermodynamical stability in the canonical ensemble}
\label{sec_TsallisC} 

In the canonical ensemble, the thermodynamical parameter is the free energy (Massieu function) $J=S-{1\over T}E$ which is explicitly given by
\begin{equation}
J=-{n\over T}\int p d^{3}{\bf r}-{1\over 2T}\int\rho\Phi d^{3}{\bf r},
\label{TsallisC1}
\end{equation}
where we have used Eqs. (\ref{sp9}) and (\ref{sp10}).  Using the
equation of state (\ref{sp6}), the second order variations of free
energy can be put in the form
\begin{equation}
\delta^{2}J=-{1\over 2T}\biggl\lbrace\int {n+1\over n}{p\over\rho^{2}}(\delta\rho)^{2} d^{3}{\bf r}+\int \delta\rho\delta\Phi d^{3}{\bf r}\biggr\rbrace.
\label{TsallisC2}
\end{equation}
As in previous papers, we introduce the function $q(r)$ defined by
\begin{equation}
\delta\rho={1\over 4\pi r^{2}}{dq\over dr},
\label{TsallisC3}
\end{equation}
and satisfying the perturbed Gauss theorem
\begin{equation}
{d\delta\Phi\over dr}={Gq\over r^{2}}.
\label{TsallisC3bis}
\end{equation}
The conservation of mass imposes $q(0)=q(R)=0$. After an integration by parts, 
we can rewrite Eq. (\ref{TsallisC2}) in the form
\begin{equation}
\delta^{2}J={1\over 2T}\int_{0}^{R}dr q\biggl\lbrack K\gamma {d\over dr}\biggl ({\rho^{\gamma-2}\over 4\pi r^{2}}{d\over dr}\biggr )+{G\over r^{2}}\biggr\rbrack q.
\label{TsallisC4}
\end{equation}
The point of marginal stability is therefore determined by the condition
\begin{equation}
K\gamma {d\over dr}\biggl ({\rho^{\gamma-2}\over 4\pi r^{2}}{dq\over dr}\biggr )+{Gq\over r^{2}}=0.
\label{TsallisC5}
\end{equation}
This is the same equation as that determining the dynamical stability
of gaseous polytropes (Chavanis 2002b). We have solved this equation
and showed that the boundary condition $q(R)=0$ returns the condition
of Eq. (\ref{tpc5}) brought by the turning point criterion. The
structure of the density perturbation profile that triggers
instability is also discussed in our previous paper.

\subsection{Generalized thermodynamical stability in the grand canonical ensemble}
\label{sec_TsallisGC} 

The thermodynamical potential in the grand canonical ensemble is $G=S-{1\over T}E+{\mu\over T}M$, i.e.
\begin{eqnarray}
G=
n z^{1/n}\int \rho d^{3}{\bf r}-{n\over T}\int p d^{3}{\bf r}
-{1\over 2T}\int\rho\Phi d^{3}{\bf r}.
\label{TsallisCG1}
\end{eqnarray}
Its second order variations are given by Eq. (\ref{TsallisC2}) like for the free energy. Only the boundary conditions are different since the mass is not conserved anymore. Introducing the function $X(r)$ defined by Eq. (\ref{gc4}), we obtain after an integration by parts
\begin{eqnarray}
\delta^{2}G={1\over 2TG^{2}}\int_{0}^{R}dr X'\biggl\lbrack {G}+K\gamma {d\over dr}\biggl ({\rho^{\gamma-2}\over 4\pi}{d\over dr}\biggr )\biggr\rbrack X'.
\label{TsallisCG2}
\end{eqnarray}
The point of marginal stability is determined by the equation 
\begin{eqnarray}
K\gamma {d\over dr}\biggl ({\rho^{\gamma-2}\over 4\pi}{X''}\biggr )+GX'=0,
\label{TsallisCG3}
\end{eqnarray}
with the boundary conditions (\ref{gc5}). Integrating once, we get
\begin{eqnarray}
K\gamma {\rho^{\gamma-2}\over 4\pi}{X''}+GX=0.
\label{TsallisCG4}
\end{eqnarray}
In terms of the dimensionless variables introduced in Sec. \ref{sec_lane}, we have to solve the problem
\begin{eqnarray}
{1\over n}\theta^{1-n}X''+X=0,
\label{TsallisCG5}
\end{eqnarray}
\begin{eqnarray}
X(0)=X'(\alpha)=0.
\label{TsallisCG6}
\end{eqnarray}
Let us introduce the differential operator
\begin{eqnarray}
{\cal L}={1\over n}\theta^{1-n}{d^{2}\over d\xi^{2}}+1.
\label{op1}
\end{eqnarray}
Using the Lane-Emden equation (\ref{lane2}), it is easy to check that 
\begin{eqnarray}
{\cal L}(\xi)=\xi,\quad {\cal L}(\xi^{2}\theta')=-{2\over n}\xi\theta,\quad {\cal L}(\xi\theta)={n-1\over n}\xi\theta.
\label{op2}
\end{eqnarray}
Therefore, the general solution of Eq. (\ref{TsallisCG5}) satisfying
$X(0)=0$ is
\begin{eqnarray}
X=\xi\biggl (\xi\theta'+{2\over n-1}\theta\biggr ).
\label{TsallisCG7}
\end{eqnarray}
The critical value of $\alpha$ at which the series of equilibria
becomes unstable in GCE is determined by the boundary condition
$X'(\alpha)=0$. Using Eq. (\ref{TsallisCG7}) and introducing the Milne
variables (\ref{lane4}), we find that this condition is equivalent to
Eq. (\ref{tpc11}) in agreement with the turning point criterion. The
density perturbation profile triggering instability can be deduced
from Eqs. (\ref{q1}) and (\ref{TsallisCG7}). We get
\begin{eqnarray}
{\delta\rho\over \rho}={3\over 4\pi}c_{1} (v_{s}-v),
\label{TsallisCG8}
\end{eqnarray}
with $v_{s}=2/(n-1)$. Equation (\ref{TsallisCG8}) coincides with the
expression found in CE. The number of nodes of the perturbation
profile can be determined by a graphical construction like in Chavanis
(2002b). First, consider the case $n=5$ which can be solved
analytically. The solution of the equation $v(\xi)=v_{s}$ is
$\xi_{(1)}=\sqrt{3}$. Since $\xi_{(1)}>\alpha=(3/5)^{1/2}$, we
conclude that the perturbation profile in GCE has no node contrary to
the equivalent situation in CE. This conclusion remains true for all
index $n\le 5$. Indeed, the critical case $\xi_{(1)}=\alpha$ would
occur for an index $n$ such that $(u_{s},v_{s})$ belongs to the
solution curve (substitute $v_{0}=v_{s}$ in Eq. (\ref{tpc11})). This
is never the case for $n\le 5$ (see Chandrasekhar 1942). Therefore,
the structure of the density profile for $n<5$ is the same as for
$n=5$ so it has no node. For $n>5$, the first mode of instability has no
node, the second mode of instability one node etc...

\subsection{Generalized thermodynamical stability in the grand microcanonical ensemble}
\label{sec_gmcTsallis}

In the grand microcanonical ensemble, the fixed parameters are the fugacity and the energy. The normalized energy is given  by (Taruya \& Sakagami 2002a)
\begin{eqnarray}
\Lambda\equiv -{ER\over GM^{2}}=-{1\over n-5}\biggl\lbrack {3\over 2}\biggl (1-{1\over v_{0}}\biggr )+{n-2\over n+1}{u_{0}\over v_{0}}\biggr \rbrack.
\label{gmcTsallis1}
\end{eqnarray}
Eliminating $\beta$ in Eq. (\ref{tpc10}) in profit of $E$, using Eqs. (\ref{tpc4}) and (\ref{gmcTsallis1}), we find that the relevant control parameter
\begin{eqnarray}
\nu\equiv -{(n\ z^{1/n})^{2(2n-3)}\over (n+1)^{4n}}{1\over Q^{4(n-1)}}G^{5}R^{7}E,
\label{gmcTsallis2}
\end{eqnarray}
is related to $\alpha$ by
\begin{eqnarray}
\nu=-u_{0}^{4}v_{0}^{6}(1-v_{0})^{2(2n-3)}\qquad \qquad\qquad \qquad\qquad \qquad\nonumber\\
 \times{1\over n-5}\biggl\lbrack {3\over 2}\biggl (1-{1\over v_{0}}\biggr )+{n-2\over n+1}{u_{0}\over v_{0}}\biggr \rbrack.\qquad \qquad
\label{gmcTsallis3}
\end{eqnarray}

In the grand microcanonical ensemble, we have to maximize the
thermodynamical potential ${\cal K}=S+{\mu\over T}M$ at fixed $E$ and
$\mu/T$. The entropy and the energy are expressed in terms of $\rho$
and $K$ by Eqs. (\ref{sp9}), (\ref{sp10}) and (\ref{sp6}). We vary the
density and use the energy constraint to determine the corresponding
variation of $K$ (which is related to the inverse temperature
$\beta$). After some algebra, we find that the second order variations
of ${\cal K}$ are given by
\begin{eqnarray}
\delta^{2}{\cal K}=  -{1\over 2}\int\delta\rho\delta\Phi d^{3}{\bf r}-{1\over 2}\gamma\int {p\over\rho^{2}}(\delta\rho)^{2}d^{3}{\bf r}\nonumber\\
-{2n\over 3(2n-3)}{1\over \int p \ d^{3}{\bf r}}\biggl\lbrack\int\biggl (\Phi+{3\gamma p\over 2\rho}\biggl )\delta\rho \ d^{3}{\bf r}\biggr \rbrack^{2}.
\label{gmcTsallis4}
\end{eqnarray}
(To simplify the formulae, we have not written the positive
proportionality factor in front of $\delta^{2}{\cal K}$). Introducing
the variable $X$ defined previously and integrating by parts, we can
put the problem in the form
\begin{eqnarray}
\delta^{2}{\cal K}=\int_{0}^{R}dr\int_{0}^{R} dr' X'(r)K(r,r')X'(r'),
\label{gmcTsallis5}
\end{eqnarray} 
with
\begin{eqnarray}
K(r,r')=\biggl\lbrace {G}+\gamma{d\over dr}\biggl ({p\over 4\pi \rho^{2}}{d\over dr}\biggr )\biggr\rbrace\delta(r-r')
-{4n\over 3(2n-3)}\nonumber\\
\times{1\over \int p \ d^{3}{\bf r}}  \biggl \lbrack \biggl (\Phi+{3\gamma p\over 2\rho}\biggr ) r\biggr\rbrack'(r)\biggl \lbrack \biggl (\Phi+{3\gamma p\over 2\rho}\biggr ) r\biggr\rbrack'(r').\qquad
\label{gmcTsallis6}
\end{eqnarray} 
The condition of marginal stability reads 
\begin{eqnarray}
{GX}+{\gamma p\over 4\pi\rho^{2}}X''=
{4n\over 3(2n-3)}{1\over \int p \ d^{3}{\bf r}} \biggl(\Phi+{3\gamma p\over 2\rho}\biggr ) r \nonumber\\
\times\int_{0}^{R} \biggl \lbrack \biggl (\Phi+{3\gamma p\over 2\rho}\biggr ) r\biggr\rbrack'   X'dr.\qquad
\label{gmcTsallis9}
\end{eqnarray} 
We now need to relate the gravitational potential $\Phi$ to the
Lane-Emden function $\theta$. Integrating the equation of hydrostatic
equilibrium for a polytropic equation of state, we
readily find that
\begin{eqnarray}
(n+1){p\over\rho}=-\Phi+{\rm Const.}
\label{wq1}
\end{eqnarray} 
The constant can be determined from the boundary condition $\Phi(R)=-GM/R$. Using the relation
\begin{eqnarray}
{GM\over R}=-(n+1)K\rho_{0}^{1/n}\alpha\theta'(\alpha),
\label{wq2}
\end{eqnarray}  
which results from Eqs. (\ref{tpc3}) and (\ref{lane3}), we find that Eq. (\ref{wq1}) can be written
\begin{equation}
\Phi=(n+1)K\rho_{0}^{1/n}(\theta(\alpha)-\theta+\alpha\theta'(\alpha)).
\label{gmcTsallis10}
\end{equation} 
Inserting this relation in Eq. (\ref{gmcTsallis9}) and introducing the dimensionless variables defined in Sec. \ref{sec_lane}, we find that
\begin{eqnarray}
X+{1\over n}\theta^{1-n}X''=\qquad\qquad\qquad\qquad\qquad\qquad \nonumber\\
{2(n+1)\xi\over \int_{0}^{\alpha}\theta^{1+n}\xi^{2}d\xi}
\biggl\lbrack \theta(\alpha)+{3-2n\over 2n}\theta+\alpha\theta'(\alpha)\biggr\rbrack\nonumber\\
\times\biggl\lbrace -{1\over 3}\int_{0}^{\alpha}\xi \theta^{n}Xd\xi+{\alpha\theta'(\alpha)+\theta(\alpha)\over 2n-3}X(\alpha)\biggr\rbrace,
\label{gmcTsallis11}
\end{eqnarray} 
where we have integrated by parts and used the identity $(\xi\theta)''=-\xi \theta^{n}$ equivalent to the Lane-Emden equation (\ref{lane2}). We are led 
therefore to solve a problem of the form
\begin{equation}
{1\over n}\theta^{1-n}X''+X=V\theta\xi+W\xi,
\label{gmcTsallis14}
\end{equation} 
\begin{equation}
X(0)=X'(\alpha)=0.
\label{gmcTsallis15}
\end{equation} 
Using the identities  (\ref{op2}), we find that the general solution of Eq. (\ref{gmcTsallis14}) satisfying $X(0)=0$ is
\begin{equation}
X=W\xi+b\biggl \lbrack \xi\theta+{1\over 2}(n-1)\xi^{2}\theta'\biggr \rbrack-{n\over 2}V\xi^{2}\theta'.
\label{gmcTsallis18}
\end{equation}
The constant $b$ is determined by the boundary condition $X'(\alpha)=0$. Substituting the resulting expression for $X(\xi)$ in Eq. (\ref{gmcTsallis11}), and using the identities of Appendix \ref{sec_id2}, we obtain after tedious algebra a condition for $\alpha$ which is equivalent to the one obtained from the turning point argument, i.e. by setting $d\nu/d\alpha=0$.     

\section{The isobaric ensemble}
\label{sec_isobaric}

We now consider the stability of isothermal spheres in contact with a
medium exerting a constant pressure $P$ on their boundary. In this
{isobaric ensemble}, the volume of the system can change unlike in
the Antonov problem. This situation was first considered by Ebert
(1955), Bonnor (1956) and McCrea (1957) who found that an isothermal
gas becomes gravitationally unstable above a critical pressure
$P_{max}$ or above a maximum compression. We shall complete these
results by supplying an entirely { analytical} derivation of the stability
criterion and by showing the equivalence between thermodynamical and
dynamical stability (Jeans problem). We shall also consider the case
of polytropic gas spheres under external pressure.

\subsection{Isothermal gas spheres under external pressure}
\label{sec_isoP}

Let us consider an isothermal gas sphere with mass $M$ at temperature
$T$ and under pressure $P$. The first principle of thermodynamics can
be written $\delta S-{1\over T}\delta E-{p\over T}\delta V=0$. In the
{\it isothermal-isobaric} ensemble (TBE), the relevant thermodynamical
potential is the Gibbs energy ${\cal G}=S-{1\over T}E-{PV\over
T}$. Its first variations satisfy $\delta {\cal G}=-E\delta({1\over
T})-V\delta({P\over T})$ so that, at statistical equilibrium, the system is in
the state that maximizes ${\cal G}$ at fixed temperature $T$ and
pressure $P$.

\begin{figure}
\centering
\includegraphics[width=8.5cm]{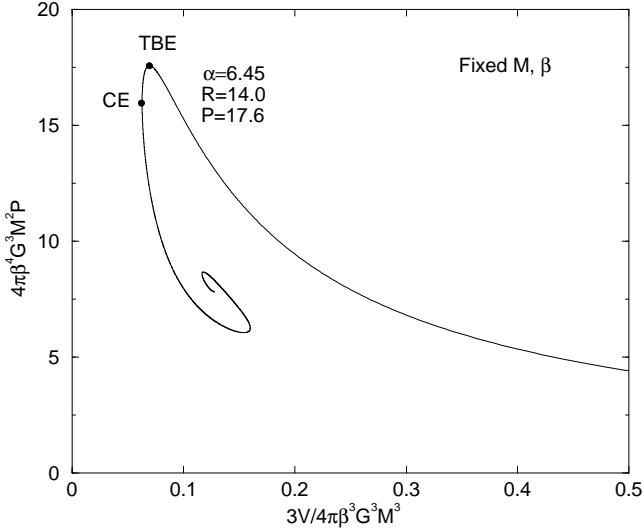}
\caption{The $P-V$ diagram for isothermal spheres. The series of equilibria becomes unstable at TBE in the isothermal-isobaric ensemble.}
\label{TBEiso}
\end{figure}

For an isothermal gas, $p={k\over m}\rho T$. Using Eq. (\ref{emden1}), the pressure on the boundary of the sphere is
\begin{eqnarray}
P={kT\over m}\rho(R)={\rho_{0}\over\beta}e^{-\psi(\alpha)}.
\label{isoP1}
\end{eqnarray}
The central density $\rho_{0}$ can be eliminated via the relation (\ref{alp}). Eliminating the radius $R$ via the relation (\ref{eq2}) and introducing the Milne variables (\ref{emden3}), we obtain after simplification 
\begin{eqnarray}
{\cal P}\equiv 4\pi G^{3}\beta^{4}M^{2}P=u_{0}v_{0}^{3}.
\label{isoP2}
\end{eqnarray}
Similarly, the normalized volume of the sphere is given by
\begin{eqnarray}
{\cal V}\equiv {3V\over 4\pi \beta^{3}G^{3}M^{3}}={1\over v_{0}^{3}}.
\label{isoP3}
\end{eqnarray}
Using the expansion of the Milne variables for $\alpha\rightarrow 0$,
\begin{eqnarray}
u_{0}=3-{1\over 5}\alpha^{2}+...,\qquad v_{0}={1\over 3}\alpha^{2}-...,
\label{isoP4}
\end{eqnarray}
we find that the first order correction to Boyle's law due to self-gravity is
\begin{eqnarray}
PV=NkT-\lambda {GM^{2}\over V^{1/3}},
\label{isoP5}
\end{eqnarray}
with $\lambda={1\over 5}({4\over 3}\pi)^{1/3}$. This formula can also be derived from the Virial theorem
\begin{eqnarray}
PV=\int p\ d^{3}{\bf r}+{1\over 3}W,
\label{isoP5bis}
\end{eqnarray}
by using the isothermal equation of state and by assuming that the density $\rho$ is uniform (in which case the
potential energy $W=-3GM^{2}/5R$). As reported by Bonnor (1956),
Eq. (\ref{isoP5}) was first suggested by Terletsky (1952) as a simple
attempt to take into account gravitational effects in the perfect gas
law. It appears here as a systematic expansion in terms of the
concentration factor $\alpha$. For large concentrations,
Eq. (\ref{isoP5}) is not valid anymore and Eqs. (\ref{isoP2})-(\ref{isoP3}) must be used instead. The $P-V$ curve is plotted in
Fig. \ref{TBEiso}. It presents a striking spiral behaviour towards the
limit point $({\cal V}_{s},{\cal P}_{s})=(1/8,8)$ corresponding to the
singular sphere. This is similar to the $E-T$ curve and other curves
studied in Sec. \ref{sec_eq}. This is interesting to note, for
historical reasons, because this diagram was discoved before the
fundamental papers of Antonov (1962) and Lynden-Bell \& Wood (1968) on
the gravothermal catastrophe. Like the minimum energy of Antonov,
there exists a maximum pressure $P_{max}=1.40\ k^{4}T^{4}/
G^{3}M^{2}m^{4}$ above which no hydrostatic equilibrium for an
isothermal gas is possible. Alternatively, this corresponds to a minimum
temperature $T_{min}=0.919\ G^{3/4}mM^{1/2}P^{1/4}/k$ for a given mass
and pressure. From the turning point criterion, the series
of equilibria becomes unstable in TBE at the point
of maximum pressure. The condition $d{\cal P}/d\alpha=0$ leads to
\begin{eqnarray}
2u_{0}-v_{0}=0.
\label{isoP6}
\end{eqnarray}
The critical value $\alpha$ is determined by the intersection between
the spiral in the $(u,v)$ plane and the straight line
(\ref{isoP6}). Its numerical value is $\alpha=6.45$ corresponding to a
density contrast of $14.0$. More compressed isothermal spheres,
ie. those for which $V\le V_{min}=0.29\ G^{3}M^{3}m^{3}/k^{3}T^{3}$
are thermodynamically unstable.

Using Eqs. (\ref{isoP2})-(\ref{isoP3}) and the identities (\ref{emden4})-(\ref{emden5}), we readily obtain
\begin{eqnarray}
{d\ln{\cal V}\over d\alpha}=-{3\over \alpha }(u_{0}-1),
\label{isoP7}
\end{eqnarray} 
and
\begin{eqnarray}
{d\ln {\cal P}\over d\alpha}={1\over \alpha}(2u_{0}-v_{0}).
\label{isoP8}
\end{eqnarray} 
Therefore, the ``compressibility'' $\kappa=-({\partial\ln P/ \partial \ln V})_{N,T}$ can be expressed as
\begin{eqnarray}
\kappa={1\over 3}{2u_{0}-v_{0}\over u_{0}-1}.
\label{isoP9}
\end{eqnarray} 
Using $u_{0}=3PV/NkT$ and $v_{0}=(4\pi/3)^{1/3}GMm/kTV^{1/3}$
according to Eqs. (\ref{isoP2}) and (\ref{isoP3}), we can check that
Eq. (\ref{isoP9}) is equivalent to Eq. (2.16) of Bonnor (1956), but it
has been obtained here in a much more direct manner. Gravitational
instability in TBE occurs when the compressibility becomes {\it
negative} passing by $\kappa=0$.

We now consider the dynamical stability of isothermal gas spheres
under external pressure and show the equivalence with the
thermodynamical criterion. This problem was studied numerically by
Yabushita (1968) but we provide here an entirely analytical
solution. In a preceding paper, we have found that
the equation of radial pulsations for a gas with an isothermal
equation of state could be written (Chavanis 2002a)
\begin{eqnarray}
{k\over m}{d\over dr}\biggl ({1\over 4\pi \rho r^{2}}{dq\over dr}\biggr )+{Gq\over Tr^{2}}={\lambda^{2}\over 4\pi\rho Tr^{2}}q,
\label{isoP10}
\end{eqnarray} 
where $\lambda$ is the growth rate of the perturbation (such that $\delta\rho\sim e^{\lambda t}$) and $q$ is defined by Eq. (\ref{TsallisC3}). In the isobaric ensemble, the boundary conditions are given by Eq. (\ref{puls4}) of Appendix \ref{sec_puls}. Introducing dimensionless variables and considering the case of marginal stability $\lambda=0$, we are led to solve the problem 
\begin{eqnarray}
{d\over d\xi}\biggl ({e^{\psi}\over\xi^{2}}{dF\over d\xi}\biggr )+{F(\xi)\over \xi^{2}}=0,
\label{isoP11}
\end{eqnarray} 
with $F(0)=0$ and
\begin{eqnarray}
{dF\over d\xi}+{d\psi\over d\xi}F=0, \qquad {\rm at}\qquad  \xi=\alpha.
\label{isoP12}
\end{eqnarray} 
Now, the function $F(\xi)$ can be expressed as (Chavanis 2002a)
\begin{eqnarray}
F(\xi)=c_{1}(\xi^{3}e^{-\psi}-\xi^{2}\psi').
\label{isoP13}
\end{eqnarray} 
Substituting this relation in the boundary condition (\ref{isoP12}), using the Emden equation (\ref{emden2}) and introducing the Milne variables (\ref{emden3}), we obtain the condition
\begin{eqnarray}
2u_{0}-v_{0}=0,
\label{isoP14}
\end{eqnarray} 
which coincides with the criterion (\ref{isoP6}) of thermodynamical
stability. The expressions of the density perturbation profile and of
the velocity profile at the point of marginal stability are given in
Chavanis (2002a). Only the value of the critical point $\alpha$
changes.  Due to the different boundary conditions, we now have
$\delta v\neq 0$ at the surface of the gaseous sphere since the sphere
can oscillate. The perturbation profile $\delta\rho/\rho$ has one node
for the first mode of instability, two nodes for the second mode
etc... The velocity profile $\delta v$ has no node for the first mode
(except at the origin), one node for the second mode etc...

\begin{figure}
\centering
\includegraphics[width=8.5cm]{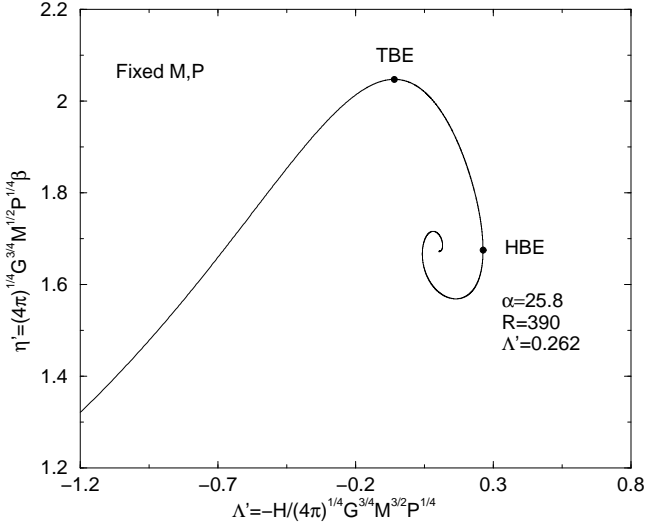}
\caption{Temperature vs enthalpy plot at fixed pressure and mass. The series of equilibria becomes unstable at HBE in the isenthalpic-isobaric ensemble.}
\label{ABE}
\end{figure}

We could also consider the situation in which the pressure $P$ and the
enthalpy $H=E+PV$ are fixed (instead of the temperature). This
corresponds to the {\it isenthalpic-isobaric} ensemble (HBE). Since
$\delta S={1\over T}\delta H-{V\over T}\delta P$, the entropy must be
a maximum at fixed $H$ and $P$ at statistical equilibrium. Eliminating
the radius $R$ in Eqs. (\ref{eq1}) and (\ref{eq2}) in profit of the
pressure $P$, we get after some algebra
\begin{eqnarray}
\Lambda'\equiv {-H\over (4\pi)^{1/4}G^{3/4}M^{3/2}P^{1/4}}={1\over (u_{0}v_{0}^{3})^{1/4}}\biggl ({3\over 2}-{4\over 3}u_{0}\biggr ),\nonumber\\
\label{isoP17}
\end{eqnarray} 
\begin{eqnarray}
\eta'\equiv (4\pi)^{1/4}G^{3/4}M^{1/2}P^{1/4}\beta= (u_{0}v_{0}^{3})^{1/4}.
\label{isoP18}
\end{eqnarray}
The $T-H$ curve in HBE is plotted in Fig. \ref{ABE}. It spirals around
the fixed point $(\Lambda'_{s},\eta'_{s})=(2^{1/4}/12,2^{3/4})$
corresponding to the singular solution. There is no equilibrium state
for $\Lambda'>0.262$. This implies a minimum enthalpy for a fixed
pressure or a maximum pressure for a fixed enthalpy. This is similar
to the Antonov (1962) criterion, the pressure playing the same role
as the box radius.  The series of equilibria becomes unstable in HBE at
the point of minimum enthalpy. The condition $d\Lambda'/d\alpha=0$
gives
\begin{eqnarray}
16u_{0}^{2}+8u_{0}v_{0}-38u_{0}+3v_{0}=0.
\label{isoP19}
\end{eqnarray} 
The critical value of $\alpha$ is $25.8$ corresponding to a density
contrast of $390$. 

The specific heat at fixed pressure
$C_{P}=(\partial H/\partial T)_{N,P}$ can be deduced from Eqs. (\ref{isoP17}) and (\ref{isoP18}). After simplification, we find
\begin{eqnarray}
C_{p}={1\over 2}Nk{16u_{0}^{2}+8u_{0}v_{0}-38u_{0}+3v_{0}\over 2u_{0}-v_{0}}.
\label{isoP20}
\end{eqnarray} 
For the ideal gas without gravity $(u_{0},v_{0})=(3,0)$ we recover the
well-known result $C_{P}={5\over 2}Nk$.  Gravitational instability
occurs in the isenthalpic-isobaric ensemble when $C_{P}=0$ passing from
negative to positive values and in the isothermal-isobaric ensemble when $C_{P}<0$.

\subsection{Polytropic gas spheres under external pressure}
\label{sec_polyP}

\begin{figure}
\centering
\includegraphics[width=8.5cm]{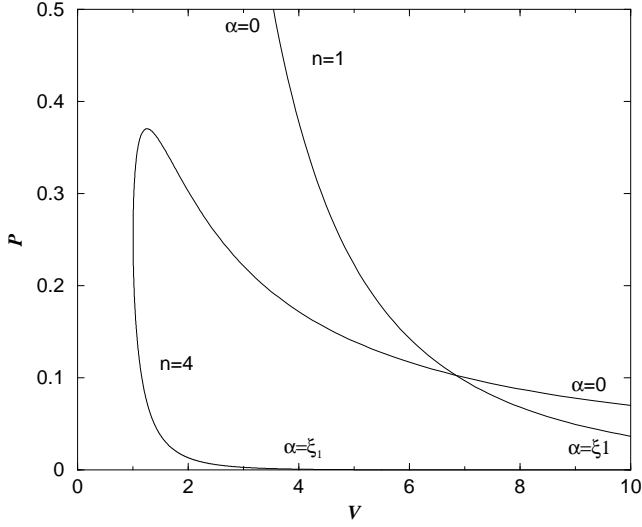}
\caption{The $P-V$ diagram for polytropes with $1\le n<3$ (specifically $n=1$) and $3<n\le 5$ (specifically $n=4$).}
\label{PVpoly1}
\end{figure}

\begin{figure}
\centering
\includegraphics[width=8.5cm]{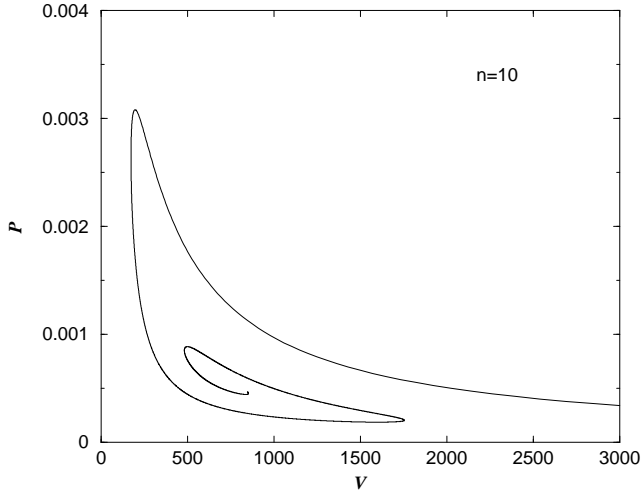}
\caption{The $P-V$ diagram for polytropes with $n>5$ (specifically $n=10$).}
\label{PVpoly2}
\end{figure}

We now consider a polytropic gas sphere of index $n$ (or $\gamma=1+1/n$) and mass $M$ under pressure $P$. Using Eqs. (\ref{sp6}) and (\ref{lane1}), the pressure on the boundary of the sphere is given by
\begin{eqnarray}
P=K\rho(R)^{\gamma}=K\rho_{0}^{1+{1\over n}}\theta(\alpha)^{n+1}.
\label{polyP1}
\end{eqnarray}
The central density $\rho_{0}$ can be eliminated via the relation (\ref{lane3}). Eliminating the radius $R$ via the relation (\ref{tpc3}) and introducing the Milne variables (\ref{lane4}), we obtain after simplification 
\begin{eqnarray}
{\cal P}\equiv {1\over K}\biggl\lbrack {4\pi G^{3}M^{2}\over K^{3}(n+1)^{3}}\biggr \rbrack^{n+1\over n-3}P=(u_{0}v_{0}^{3})^{n+1\over n-3}.
\label{polyP2}
\end{eqnarray}
Similarly, the normalized volume of the sphere is given by
\begin{eqnarray}
{\cal V}\equiv {3V}\biggl \lbrack {K^{3}(1+n)^{3}\over 4\pi G^{3}M^{3(n-1)\over n}}\biggr\rbrack^{n\over n-3}={1\over (u_{0}v_{0}^{n})^{3\over n-3}}. 
\label{polyP3}
\end{eqnarray}
Using the expansion of the Milne variables for $\alpha\rightarrow 0$,
\begin{eqnarray}
u_{0}=3-{n\over 5}\alpha^{2}+...,\qquad v_{0}={1\over 3}\alpha^{2}-...,
\label{polyP4}
\end{eqnarray}
we find that the first order correction to the classical adiabatic law due to self-gravity is
\begin{eqnarray}
PV^{\gamma}=KM^{\gamma}-\lambda {GM^{2}\over V^{n-3\over 3n}},
\label{polyP5}
\end{eqnarray}
Equation (\ref{polyP5}) can also be obtained from the Virial theorem
(\ref{isoP5bis}) as before. The $P-V$ curve is plotted in
Figs. \ref{PVpoly1} and \ref{PVpoly2} for different values of the
polytropic index $n$. The volume is extremum at points $\alpha$ such
that $d{\cal V}/d\alpha=0$.  This leads to the condition
$u_{0}=u_{s}$. The pressure is extremum at
points $\alpha$ such that $d{\cal P}/d\alpha=0$. This leads to the
condition
\begin{eqnarray}
2u_{0}+(3-n)v_{0}=0.
\label{polyP6}
\end{eqnarray}
The number of extrema can be obtained by a graphical construction in the $(u,v)$ plane like in Chavanis (2002b). For $n<3$, there is no extremum for ${\cal P}$ and ${\cal V}$. The pressure is a decreasing function of the volume and vanishes for a maximum volume $V_{max}$ corresponding to the complete polytrope (such that $\alpha=\xi_{1}$). For $n=1$, the solution of the Lane-Emden equation is $\theta=\sin\xi/\xi$ for $\xi\le \xi_{1}=\pi$ and we explicitly obtain
\begin{eqnarray}
{\cal P}={\sin^{2}\alpha\over\alpha^{2}(\sin\alpha-\alpha\cos\alpha)^{2}},\qquad {\cal V}=\alpha^{3},
\label{polyP6a}
\end{eqnarray}
for $\alpha\le\pi$. For $3<n\le 5$, there is one extremum for ${\cal P}$ and ${\cal V}$. This implies the existence of a maximum pressure $P_{max}$ and a minimum volume $V_{min}$. In particular, for $n=5$, the Lane-Emden equation can be solved analytically and we get
\begin{eqnarray}
{\cal P}={\alpha^{18}\over 3^{6}(1+{1\over 3}\alpha^{2})^{12}},\qquad {\cal V}={3^{6}\over\alpha^{15}}\biggl (1+{1\over 3}\alpha^{2}\biggr )^{9}.
\label{polyP6b}
\end{eqnarray}
The pressure is maximum for $\alpha=3$ and the volume is minimum for
$\alpha=\sqrt{15}$. For $n>5$ there is an infinity of
extrema for ${\cal P}$ and ${\cal V}$. The $P-V$ curve spirals towards
the limit point $({\cal V}_{s},{\cal P}_{s})$ corresponding to the
singular sphere. From turning point arguments, the series of
equilibria becomes thermodynamically unstable at the point of maximum
pressure (for $n>3$). Polytropes with $n<3$ are always stable in the
isobaric ensemble.

Using Eqs. (\ref{polyP2}), (\ref{polyP3}) and the identities (\ref{lane5}) and (\ref{lane6}), we readily obtain
\begin{eqnarray}
{d\ln{\cal V}\over d\alpha}=-{3\over \alpha}\biggl ({n-1\over n-3}u_{0}-1\biggr ),
\label{polyP7}
\end{eqnarray} 
and
\begin{eqnarray}
{d\ln{\cal P}\over d\alpha}={n+1\over (n-3)\alpha}\bigl ((3-n)v_{0}+2u_{0}\bigr).
\label{polyP8}
\end{eqnarray} 
Therefore, the ``compressibility'' $\kappa=({\partial\ln P/ \partial \ln V})_{N,K}$ can be expressed as
\begin{eqnarray}
\kappa=-{1\over 3}(n+1){2u_{0}+(3-n)v_{0}\over (n-1)u_{0}-n+3}.
\label{polyP9}
\end{eqnarray}

We now consider the dynamical stability of polytropic gas spheres
under external pressure. The equation
of radial pulsations for a gas with a polytropic equation of state
can be written (Chavanis 2002b)
\begin{eqnarray}
K\gamma{d\over dr}\biggl ({\rho^{\gamma-2}\over 4\pi r^{2}}{dq\over dr}\biggr )+{Gq\over r^{2}}={\lambda^{2}\over 4\pi\rho r^{2}}q.
\label{polyP10}
\end{eqnarray} 
In the isobaric ensemble, the boundary conditions are given by
Eq. (\ref{puls4}) of Appendix \ref{sec_puls}. Introducing
dimensionless variables and considering the case of marginal stability
$\lambda=0$, we are led to solve the problem
\begin{eqnarray}
{d\over d\xi}\biggl ({\theta^{1-n}\over\xi^{2}}{dF\over d\xi}\biggr )+{nF(\xi)\over \xi^{2}}=0,
\label{polyP11}
\end{eqnarray} 
with $F(0)=0$ and
\begin{eqnarray}
\theta{dF\over d\xi}-n{d\theta\over d\xi}F=0, \qquad {\rm at}\qquad  
\xi=\alpha.
\label{polyP12}
\end{eqnarray} 
Now, the function $F(\xi)$ can be expressed as (Chavanis 2002b)
\begin{eqnarray}
F(\xi)=c_{1}\biggl (\xi^{3}\theta^{n}+{n-3\over n-1}\xi^{2}\theta'\biggr ).
\label{polyP13}
\end{eqnarray} 
Substituting this relation in the boundary condition (\ref{polyP12}), using the Lane-Emden equation (\ref{lane2}) and introducing the Milne variables (\ref{lane4}), we obtain the condition
\begin{eqnarray}
2u_{0}+(3-n)v_{0}=0,
\label{polyP14}
\end{eqnarray} 
which coincides with the criterion (\ref{polyP6}) of thermodynamical
stability. The expressions of the density perturbation profile and of
the velocity profile at the point of marginal stability are given in
Chavanis (2002b). Only the value of the critical point $\alpha$
changes. For $3<n\le 5$, the density perturbation profile has one node
and the velocity profile no node (except at the origin).  For $n>5$,
the perturbation profile $\delta\rho/\rho$ has one node for the first
mode of instability, two nodes for the second mode etc... The velocity
profile $\delta v$ has no node for the first mode (except at the
origin), one node for the second mode etc...

\section{Stability of galaxies and stars}
\label{sec_st}

In this section, we discuss in detail the relation between the
dynamical and the thermodynamical stability of stellar systems and
gaseous stars following our previous investigations (Chavanis
2002a,b). This discussion should clarify the notion of generalized
thermodynamics introduced by Tsallis (1988).

\subsection{Collisional vs collisionless relaxation}
\label{sec_collness}

Basically, a stellar system is a collection of $N$ stars in
gravitational interaction. Due to the development of encounters
between stars, this system is expected to relax towards a statistical
equilibrium state. The relaxation time can be estimated by
$t_{relax}\sim (N/\ln N)t_{D}$ where $t_{D}$ is the dynamical time,
i.e. the time it takes a star to cross the system (Binney \& Tremaine
1987). For globular clusters, $N\sim 10^{5}$, $t_{D}\sim 10^{5}$ years
so that the relaxation time is of the same order as their age ($\sim
10^{10}$ years). Therefore, globular clusters form a {\it discret
Hamiltonian system} which can be studied by ordinary methods of
statistical mechanics. In particular, it can be shown rigorously that
the Boltzmann entropy $S_{B}[f]$ is the correct form of entropy for
these systems (in a suitable thermodynamic limit) although they are
non-extensive and non-additive. The absence of entropy maximum in an
unbounded domain simply reflects the natural tendency of a stellar
system to evaporate under the effect of encounters. However,
evaporation is a slow process so that a globular cluster passes by a
succession of quasi-equilibrium states corresponding to truncated
isothermals (see Sec. \ref{sec_field}). It should be emphasized that 
statistical mechanics is based on the postulate that the evolution is 
ergodic. Now, the kinetic theory of self-gravitating systems is
not well-understood (Kandrup 1981). Indeed, it is possible to prove a
H-theorem for the Boltzmann entropy only if simplifying assumptions
are implemented (Markovian approximation, local approximation,
regularization of logarithmic divergences,...). It would be of
interest to determine more precisely the influence of non-ideal
effects on the relaxation of stars. However, these effects are not
expected to induce a strong deviation to Boltzmann's law. In
particular, the diffusion of stars is normal or slightly anomalous,
i.e. polluted by logarithmic corrections (Lee 1968). Hence, Tsallis
entropies, which often arise in the case of anomalous diffusion, do
not seem to be justified in the context of ``collisional'' stellar
systems. However, there are many effects that can induce a lack of
ergodicity and a deviation to Boltzmann's law. To a large extent, the
dynamics of self-gravitating systems remains misunderstood and
additional work (using numerical simulations) is necessary to clarify
the process of collisional relaxation.
 
For elliptical galaxies, on the other hand, the relaxation time due to
close encounters is larger than their age by about ten orders of
magnitude ($N\sim 10^{11}$, $t_{D}\sim 10^{8}$ years, age $\sim
10^{10}$ years). Therefore, their dynamics is essentially
encounterless and described by the self-consistent Vlasov-Poisson
system. Elliptical galaxies form therefore a {\it continuous
Hamiltonian system}. Now, due to a complicated mixing process in phase
space, the Vlasov-Poisson system can achieve a metaequilibrium state
(on a coarse-grained scale) on a very short timescale. A statistical
theory appropriate to this {\it violent relaxation} was developed by
Lynden-Bell in 1967 (see a short review in Chavanis 2002f). For
collisionless stellar systems, the infinite mass problem is solved by
realizing that violent relaxation is {\it incomplete} so that the
galaxies are more confined that one would expect on the basis of
statistical considerations.

Collisionless and collisional relaxations are therefore two distinct
processes in the life of a self-gravitating system and they must be
studied by different types of statistical mechanics.  Note that a
globular cluster first experiences a collisionless relaxation leading
to a virialized state and then evolves more slowly towards another
equilibrium state due to stellar encounters.  We can discuss these two
successive equilibrium states in a more formal manner. Let us consider
a fixed interval of time $t$ and let the number of particles
$N\rightarrow +\infty$. In that case, the system is rigorously
described by the Vlasov equation, and a metaequilibrium state is
achieved on a timescale independant on $N$ as a result of violent
relaxation. Alternatively, if we fix $N$ and let $t\rightarrow
+\infty$, the system is expected to relax to an equilibrium state
resulting from a collisional evolution. These two equilibrium states
are of course physically distinct. This implies that the order of the
limits $N\rightarrow +\infty$ and $t\rightarrow +\infty$ is not
interchangeable for self-gravitating systems (Chavanis 2002e).

\subsection{The Vlasov-Poisson system}
\label{sec_vp}

In Secs. \ref{sec_thermo}-\ref{sec_field}, we have implicitly
considered the case of collisional stellar systems, such as globular
clusters, relaxing via two-body encounters. We now turn to the case of
collisionless stellar systems, such as elliptical galaxies, described
by the Vlasov equation
\begin{equation}
\label{vlasov}
{\partial f\over\partial t}+{\bf v}{\partial f\over\partial {\bf r}}+{\bf F}{\partial f\over\partial {\bf v}}=0,
\end{equation}
coupled with the Poisson equation (\ref{ens3}). The Vlasov equation (\ref{vlasov}) simply states
that, in the absence of encounters, the distribution function $f$ is
conserved by the flow in phase space.  This equation conserves the usual invariants (\ref{ens1}) and (\ref{ens2}) but also
an additional class of invariants called the Casimirs
\begin{equation}
\label{casimirs}
I_{h}=\int h(f)d^{3}{\bf r}d^{3}{\bf v},
\end{equation}
where $h$ is any continuous function of $f$. This infinite set of
additional constraints is due to the collisionless nature of the
evolution and results from the conservation of $f$ and the
incompressibility of the flow in phase space. As we shall see, they
play a crucial role in the statistical mechanics of violent
relaxation. We note that the conservation of the Casimirs is equivalent to the conservation of the moments 
\begin{equation}
\label{moments}
M_{n}=\int f^{n}d^{3}{\bf r}d^{3}{\bf v}.
\end{equation}

It is well-known that {\it any} distribution function of the form
$f=f(\epsilon)$, where $\epsilon={v^{2}\over 2}+\Phi$ is the
individual energy of a star, is a stationary solution of the Vlasov
equation. This is a particular case of the Jeans theorem (Binney \&
Tremaine 1987). We shall assume furthermore that $f'(\epsilon)<0$ so
that high energy stars are less probable than low energy stars. Any
such distribution function can be obtained by extremizing a functional
of the form
\begin{eqnarray}
{S}[f]=-\int C(f)d^{3}{\bf r}\ d^{3}{\bf v},
\label{st0}
\end{eqnarray}
at fixed mass $M$ and energy $E$, where $C(f)$ is a convex function,
i.e. $C''(f)\ge 0$. Furthermore, {\it maxima} of $S[f]$ at fixed mass
and energy are dynamically stable with respect to the Vlasov-Poisson
system (Ipser 1974, Ipser \& Horwitz 1979). This is a sufficient
condition of {\it nonlinear} stability. We are tempted to believe that
it is also a necessary condition of nonlinear stability although this
seems to go against the theorem of Doremus et al (1971) which states
that all distribution functions with $f'(\epsilon)<0$ are dynamically
stable (even if they are minima or saddle points of $S$ at fixed $E$
and $M$).  In fact, this criterion is only a condition of linear
stability (Binney \& Tremaine 1987) and it may not be valid for box
confined models (its general applicability has also been
criticized). In the following, we shall consider that $f=f(\epsilon)$
is dynamically stable ``in a strong sense'' if and only if it is a
maximum of $S[f]$ at fixed $E$ and $M$. This is equivalent to
$f=f(\epsilon)$ being a minimum of $E$ at fixed $S$ and $M$ (Ipser \&
Horwitz 1979). It can be shown that stellar systems satisfying this
requirement are necessarily spherically symmetric.

Introducing appropriate Lagrange multipliers and writing the
variational principle in the form
\begin{eqnarray}
\delta S-\beta\delta E-\alpha\delta M=0,
\label{st0a}
\end{eqnarray}
one finds that
\begin{eqnarray}
C'(f)=-\beta\epsilon-\alpha.
\label{st1}
\end{eqnarray}
Since $C'$ is a monotonically increasing function of $f$, we can inverse this relation to obtain 
\begin{eqnarray}
f=F(\beta\epsilon+\alpha),
\label{st2}
\end{eqnarray}
where $F(x)=(C')^{-1}(-x)$. From the identity
\begin{eqnarray}
f'(\epsilon)=-\beta/C''(f),
\label{st2a}
\end{eqnarray}
resulting from Eq. (\ref{st1}), $f(\epsilon)$ is a monotonically decreasing function of energy if $\beta>0$. 

We note also that for each stellar systems with $f=f(\epsilon)$, there exist a corresponding barotropic gas with the same equilibrium density distribution. Indeed $\rho=\int fd^{3}{\bf v}=\rho(\Phi)$, $p={1\over 3}\int fv^{2}d^{3}{\bf v}=p(\Phi)$, hence $p=p(\rho)$. Writing explicitly  the pressure in the form
\begin{eqnarray}
p={1\over 3}\int_{\Phi}^{+\infty}F(\beta\epsilon+\alpha)4\pi \lbrack 2(\epsilon-\Phi)\rbrack^{3/2}d\epsilon,
\label{st3}
\end{eqnarray}
and taking its gradient, we obtain the condition of hydrostatic equilibrium 
\begin{eqnarray}
\nabla p=-\rho\nabla\Phi.
\label{st4}
\end{eqnarray}
Therefore, the Jeans equations of stellar dynamics reduce in these cases of spherical systems with isotropic pressure to the equation of hydrostatics for a barotropic gas.

\subsection{Violent relaxation of collisionless stellar systems}
\label{sec_vr}

The Vlasov equation does not select a universal form of entropy
$S[f]$, unlike the Boltzmann equation for example.  Therefore, the
notion of equilibrium is relatively subtle. When coupled to the
Poisson equation, the Vlasov equation develops very complex filaments
as a result of a mixing process in phase space.  In this sense, the
fine-grained distribution function $f({\bf r},{\bf v},t)$ will never
converge towards a stationary solution $f({\bf r},{\bf v})$. This is
consistent with the fact that the functionals $S[f]$ are rigorously
conserved by the flow (they are particular Casimirs). However, if we
introduce a coarse-graining procedure, the coarse-grained distribution
function $\overline{f}({\bf r},{\bf v},t)$ will relax towards a
metaequilibrium state $\overline{f}({\bf r},{\bf v})$. This
collisionless relaxation is called {\it violent relaxation}
(Lynden-Bell 1967) or chaotic mixing.  It can be shown that
$S[\overline{f}]$ calculated with the coarse-grained distribution
function at time $t>0$ is larger than $S[f]$ calculated with the
fine-grained distribution function at $t=0$ (a monotonic increase of
$H$ is not implied).  This is similar to the $H$-theorem in kinetic
theory except that $S[{f}]$ is not necessarily the Boltzmann
entropy. For that reason, $S[{f}]$ is sometimes called a H-function
(Tremaine et al 1986). Boltzmann and Tsallis functionals $S_{B}[f]$
and $S_{q}[f]$ are particular H-functions corresponding to isothermal
stellar systems and stellar polytropes.

It is important to note, furthermore, that among the infinite set of
integrals conserved by the Vlasov equation, some are more robust than
others.  For example, the moments $M^{c.g.}_{n}[\overline{f}]=\int
\overline{f}^{n}d^{3}{\bf r}d^{3}{\bf v}$  calculated with the coarse-grained distribution function vary with time for $n>1$ since
$\overline{f^{n}}\neq \overline{f}^{n}$. By contrast, the mass and the
energy are approximately conserved on the coarse-grained scale.  We
shall therefore classify the collisionless invariants in two groups:
the energy $E$ and the mass $M$ will be called {\it robust integrals}
since they are conserved by the coarse-grained dynamics while the
moments $M_{n}$ with $n> 1$ will be called {\it fragile integrals}
since they vary under the operation of coarse-graining. Note that, in
practice, the moments $M_{n>1}$ are also altered at the fine-grained
scale because, intrinsically, the system is not continuous but made of
stars. Therefore, the graininess of the medium can break the
conservation of some moments, in general those of high order, even if
we are in a regime where the evolution is essentially encounterless.

\subsection{Lynden-Bell's theory of violent relaxation}
\label{sec_lb}

Now, the question of fundamental interest is the following: given some
initial condition, can we {\it predict} the equilibrium distribution
function that the system will eventually achieve?  Lynden-Bell (1967)
tried to answer this question by using statistical mechanics
arguments. He introduced the density probability $\rho({\bf r},{\bf
v},\eta)$ of finding the value of the distribution function $\eta$ in
$({\bf r},{\bf v})$ at equilibrium.  Then, using a combinatorial
analysis he showed that the relevant measure is the Boltzmann entropy
\begin{equation}
\label{qlb1}
S\lbrack \rho\rbrack=-\int \rho\ln\rho d^{3}{\bf r}d^{3}{\bf v}d\eta,
\end{equation}
which is a functional of $\rho$. This mixing entropy has been
justified rigorously by Michel \& Robert (1994), using the concept of
Young measures. Therefore, although the system is non-extensive, the
Boltzmann entropy is the correct measure. Assuming ergodicity (which
may not be realized in practice, see Sec. \ref{sec_ir}) the most
probable equilibrium state, i.e. most mixed state, is obtained by
maximizing $S\lbrack
\rho\rbrack$ taking into account all the constraints of the dynamics. The optimal equilibrium state can be written
\begin{equation}
\label{qlb2}
\rho({\bf r},{\bf v},\eta)={1\over Z}\chi(\eta)e^{-(\beta\epsilon+\alpha)\eta},
\end{equation}
where $\chi(\eta)\equiv {\rm exp}(-\sum_{n>1}\alpha_{n}\eta^{n})$ accounts for the conservation of the fragile moments $M_{n>1}=\int \rho\eta^{n}d\eta d^{3}{\bf r}d^{3}{\bf v}$ and $\alpha$, $\beta$ are the usual Lagrange multipliers for $M$ and $E$ (robust integrals). The distinction between these two types of constraints is important (see below) and was not explicitly made by Lynden-Bell (1967). The ``partition function''
\begin{equation}
\label{part}
Z=\int_{0}^{+\infty}\chi(\eta)\ e^{-(\beta\epsilon+\alpha)\eta}d\eta,
\end{equation}
is determined by the local normalization condition $\int \rho
d\eta=1$. The condition of thermodynamical stability reads 
\begin{eqnarray}
\label{thermostab}
\delta^{2}J\equiv -{1\over 2}\int {(\delta\rho)^{2}\over \rho}d^{3}{\bf r}d^{3}{\bf v}d\eta-{1\over 2}\beta\int\delta\Phi\delta\overline{f}d^{3}{\bf r}d^{3}{\bf v}\le 0,\nonumber\\
\end{eqnarray}
for all variation $\delta\rho({\bf r},{\bf v},\eta)$ which does not
change the constraints to first order. The equilibrium coarse-grained
distribution function $\overline{f}\equiv \int \rho \eta d\eta$ can
be expressed as
\begin{equation}
\label{lb3}
\overline{f}=-{1\over \beta}{\partial \ln Z\over\partial \epsilon}=F(\beta\epsilon+\alpha)=f(\epsilon).
\end{equation}
Taking the derivative of Eq. (\ref{lb3}), it is easy to show that
\begin{equation}
\label{lb4}
\overline{f}'(\epsilon)=-\beta f_{2}, \qquad f_{2}\equiv \int \rho (\eta-\overline{f})^{2}
d\eta>0,
\end{equation} 
where $f_{2}$ is the centered local variance of the distribution
$\rho({\bf r},{\bf v},\eta)$. The integrability condition $\int f
d^{3}{\bf v}<\infty$ requires that the temperature is positive
($\beta>0$). Then, from Eq. (\ref{lb4}), we deduce that $f$ is a
monotonically decreasing function of $\epsilon$. Therefore, from
Sec. \ref{sec_vp}, we know that for each function $f(\epsilon)$ of the
form (\ref{lb3}) there exists a H-function $S[\overline{f}]$ whose
extremization at fixed mass and energy returns Eq. (\ref{lb3}).
Furthermore, it can be shown that the condition (\ref{thermostab}) of
thermodynamical stability (maximum of $S[\rho]$ at fixed $E$ and
$M_{n}$) implies that $\overline{f}$ is a nonlinearly stable solution
of the Vlasov-Poisson system. Therefore, it is a {\it maximum} of
$S[\overline{f}]$ at fixed mass and energy.  This functional depends
on the initial conditions through the Lagrange multipliers
$\alpha_{n>1}$, or equivalently through the function $\chi(\eta)$,
which are determined by the fragile moments $M_{n>1}$. Therefore,
$S[\overline{f}]$ is {\it non-universal} and can take a wide variety
of forms.  In general, $S[\overline{f}]$ differs from the ordinary
Boltzmann entropy $S_{B}[\overline{f}]$ derived in the context of
collisional stellar systems. This is due to the additional constraints
(Casimir invariants) brought by the Vlasov equation. For a given
initial condition, the H-function $S[\overline{f}]$ maximized by the
system at statistical equilibrium is uniquely determined by the
statistical theory according to Eqs. (\ref{st1}) and (\ref{lb3}).

In general, $S[\overline{f}]$ is not an entropy, unlike $S[\rho]$, as
it cannot be directly obtained from a combinatorial analysis.  There
is a case, however, where $S[\overline{f}]$ is an entropy. If the
initial condition is made of patches with value $f_{0}=\eta_{0}$ and
$f_{0}=0$ (two-levels approximation), then the entropy $S[\rho]$ is
equivalent to the Fermi-Dirac entropy $S_{F.D.}[\overline{f}]=-\int
\lbrack \overline{f}/\eta_{0}\ln \overline{f}/\eta_{0}+(1- \overline{f}/\eta_{0})\ln (1-\overline{f}/\eta_{0})\rbrack d^{3}{\bf r}d^{3}{\bf v}$ and Lynden-Bell's distribution
(\ref{lb3}) coincides with the Fermi-Dirac distribution (Lynden-Bell
1967, Chavanis \& Sommeria 1998). Moreover, all the moments can be
expressed in terms of the mass alone as $M_{n}=\eta_{0}^{n}M$. In the
non-degenerate (or dilute) limit $\overline{f}\ll \eta_{0}$, the
Fermi-Dirac entropy reduces to the Boltzmann entropy
$S_{B}[\overline{f}]=-\int \overline{f}\ln
\overline{f} d^{3}{\bf r}d^{3}{\bf v}$. Quite generally, in the dilute limit, the equilibrium distribution is a sum of isothermal distributions. 

We can note some general predictions of the statistical theory, which
are valid for any initial condition $f_{0}({\bf r},{\bf v})$. (i) The
predicted distribution function is a function
$\overline{f}=f(\epsilon)$ of the stellar energy alone in the
non-rotating case and a function $\overline{f}=f(\epsilon_{J})$ of the
Jacobi energy $\epsilon_{J}=\epsilon-{\bf \Omega}\cdot ({\bf r}\times
{\bf v})$ alone in the rotating case. (ii) $f'(\epsilon)<0$ so that
high energy stars are less probable than low energy stars. (iii)
$\overline{f}(\epsilon)$ is {\it strictly} positive for all energies
(it never vanishes). (iv) $\overline{f}(\epsilon)$ is always smaller
than the maximum value $f_{0}^{max}$ of the initial condition.  (v)
According to Eqs. (\ref{lb4}) and (\ref{st2a}), we have the general
relation $f_{2}={1/C''(\overline{f})}$. (vi) For a box-confined
system, there always exist a {\it global} maximum of
entropy. Therefore, contrary to collisional stellar systems, it is not
necessary to introduce a small-scale cut-off to make the
thermodynamical approach rigorous. This is due to the Liouville
theorem which puts an upper bound on the value of the distribution
function like the Pauli exclusion principle in quantum mechanics (see
Chavanis et al 1996, Chavanis \& Sommeria 1998).

\subsection{Incomplete relaxation}
\label{sec_ir}

The prediction of Lynden-Bell's statistical theory crucially depends
on the initial conditions. This is because we need to know the value
of the moments $M_{n>1}$ which can only be deduced from the
fine-grained distribution function at $t=0$ (say). Indeed, these
moments are fragile, or microscopic, constraints which cannot be
mesured from the coarse-grained flow at $t>0$ since they are altered
by the coarse-graining procedure. This is a specificity of continuous
Hamiltonian systems which contrasts with ordinary (discrete)
Hamiltonian systems for which the constraints are robust, or
{macroscopic}, and can be evaluated at any time (as the mass $M$ and
the energy $E$). Now, Lynden-Bell (1967) gives arguments according to
which elliptical galaxies should be non-degenerate (except possibly in
the nucleus). In that case, the dilute limit of his theory applies in
the main body of the galaxy and leads to an isothermal
distribution. The theoretical justification of an isothermal
distribution for collisionless stellar systems was considered as a
triumph in the 1960's. Indeed, Lynden-Bell's statistical theory of
violent relaxation could explain the observed isothermal core of
elliptical galaxies without recourse to collisions that operate on a
much longer timescale.

However, it is easy to see that the statistical theory of violent
relaxation cannot describe the halo of elliptical galaxies. Indeed, at
large distances, the density decreases as $r^{-2}$ resulting in the
infinite mass problem. Lynden-Bell (1967) has related this
mathematical difficulty to the physical problem of {\it incomplete
relaxation} and argued that his distribution function has to be
modified at high energies. Indeed, the relaxation is effective only in
a finite region of space and persists only for a finite period of
time. Therefore, there is no reason to maximize entropy in the whole
available space. The ergodic hypothesis which sustains the statistical
theory applies only in a restricted domain of space, in a sort of
``maximum entropy bubble'', surrounded by an unmixed region which is
only poorly sampled by the system. Accordingly, the physical picture
that emerges is the following: during violent relaxation, the system
has the {\it tendency} to reach the most mixed state described by the
distribution (\ref{lb3}). However, as it approaches equilibrium, the
mixing becomes less and less efficient and the system settles on a
state which is not the most mixed state.

How can we take into account incomplete relaxation? A first
possibility is to confine the system artificially inside a box, where
the box delimits the typical region in which the statistical theory
applies (Chavanis \& Sommeria 1998). Another possibility is to use a
parametrization of the coarse-grained dynamics of collisionless
stellar systems in the form of relaxation equations involving a space
and time dependant diffusion coefficient related to the fine-grained
fluctuations of the distribution function (Chavanis et al 1996,
Chavanis 1998). This space and time dependant diffusion coefficient
can {\it freeze} the system in a ``maximum entropy bubble'' and lead
to the formation of self-confined stellar systems. Finally, a third
possibility is to construct a model of incomplete violent relaxation
which provides a depletion of stars in the high energy tail of the
distribution function. If the system is isolated, as most elliptical
galaxies are, Hjorth \& Madsen (1993) have proposed a dynamical
scenario attempting to take into account incomplete relaxation in the
halo. They introduce a two-step process: (i) in a first step, they
assume that violent relaxation proceeds to completion in a {\it
finite} spatial region, of radius $r_{max}$, which represents roughly
the core of the galaxy (where the fluctuations are important). At this
stage, the escape energy represents no special threshold so that
negative as well as positive energy states are populated in that
region. (ii) After the relaxation process is over, positive energy
particles leave the system and particles with
$\Phi(r_{max})<\epsilon<0$ move in orbits beyond $r_{max}$, thereby
changing the distribution function to something significantly
`thinner' than a Boltzmann distribution. The crucial point to realize
is that the differential energy distribution $N(\epsilon)$, where
$N(\epsilon)d\epsilon$ is the number of stars with energy between
$\epsilon$ and $\epsilon+d\epsilon$, will be discontinuous at the
escape energy $\epsilon=0$ since there is a finite number of particles
within $r_{max}$ after the relaxation process. It can be shown that
this discontinuity implies necessarily that $f(\epsilon)\sim
(-\epsilon)^{5/2}$ for $\epsilon\rightarrow 0^{-}$ (Jaffe
1987). Therefore, the distribution function consistent with this
scenario is (Hjorth \& Madsen 1993)
\begin{eqnarray}
\overline{f}=\Biggl\lbrace \begin{array}{cc}
A e^{-\beta\eta_{0} \epsilon} & \Phi_{0}\le \epsilon< \epsilon', \\
B \ (-\epsilon)^{5/2} & \epsilon'\le\epsilon<0, \\
0  & \epsilon\ge 0.
\end{array}
\label{hm}
\end{eqnarray}  
This model corresponds formally to a composite configuration with an
isothermal core in which violent relaxation is efficient and a
polytropic halo (with index $n=4$) which is only partially relaxed (in
reality, there is a small deviation to the ideal value $n=4$ because
the potential created by the core is not exactly Keplerian; see Hjorth
\& Madsen 1993). For this model, the density decreases as $r^{-4}$ like
in elliptical galaxies. Furthermore, this model can reproduce de
Vaucouleur's $R^{1/4}$ law for the surface brightness of ellipticals
(Binney \& Tremaine 1987). Alternatively, if the system is subject to
tidal forces (this may be the case for dark matter halos), one
can propose an extended Michie-King model of the form (Chavanis 1998)
\begin{eqnarray}
\overline{f}=\Biggl\lbrace \begin{array}{cc}
\eta_{0} {e^{-\beta\eta_{0}\epsilon}-e^{-\beta\eta_{0}\epsilon_{m}}\over \lambda+e^{-\beta\eta_{0}\epsilon}}   & \epsilon< \epsilon_{m}, \\
0  & \epsilon\ge \epsilon_{m},
\end{array}
\label{chav}
\end{eqnarray}  
which  takes into account the specificities of Lynden-Bell's distribution
function (in particular the exclusion principle associated with
Liouville's theorem) and leads to a confinement of the density
at large distances (i.e. the density drops to zero at a finite
distance identified with the tidal radius).

\subsection{Tsallis entropies}
\label{sec_Tcolness}

The statistical theory of Lynden-Bell encounters some problems because
the distribution function $f(\epsilon)$ never vanishes. Therefore, we
have to invoke incomplete relaxation and introduce dynamical
constraints in order to explain the observation of confined galaxies
and solve the infinite mass problem. This is probably the correct
approach to the problem. Another line of thought, defended by Tsallis
and co-workers, is to change the form of entropy in order to obtain a
distribution function that vanishes at some maximum energy. Therefore,
Tsallis generalized thermodynamics is essentially an attempt to take
into account incomplete mixing and lack of ergodicity in complicated
systems. If we assume that the system is classically described by the
Boltzmann entropy $S_{B}[f]$ (which is true for collisionless stellar
systems only in a coarse-grained sense and in a dilute limit), then
Tsallis proposes to replace $S_{B}[f]$ by a $q$-entropy
$S_{q}[f]$. The parameter $q$ measures the efficiency of mixing ($q=1$
if the system mixes well) but is not determined by the theory.

This approach is attractive but the justification given by Tsallis and
co-workers is misleading. Fundamentally, ordinary thermodynamics (in
the sense of Lynden-Bell ) is the correct approach to the problem.
The Boltzmann entropy $S[\rho]$ has a clear physical meaning as it is
proportional to the logarithm of the disorder, where the disorder is
equal to the number of microstates consistent with a given
macrostate. Of course, if the system does not mix well, statistical
mechanics loses its power of prediction. However, the state resulting
from incomplete violent relaxation is always a nonlinearly stable
stationary solution of the Vlasov equation on a coarse-grained scale (i.e., for
$\overline{f}$). Therefore, if $\overline{f}=f(\epsilon)$, it
maximizes a $H$-function $S[\overline{f}]=-\int
C(\overline{f})d^{3}{\bf r}d^{3}{\bf v}$ at fixed mass $M$ and energy
$E$. The function $C(f)$ is influenced
by thermodynamics (mixing) but it cannot be determined by pure
thermodynamical arguments because of incomplete relaxation.

The $H$-function $S[{f}]$ selected by the system depends both on the
initial condition and on the type of mixing.  Tsallis entropy
$S_{q}[{f}]$ is just a particular H-function leading to confined
structures (stellar polytropes). Now, it is well-known that stellar
polytropes give a poor fit of elliptical galaxies. In particular, they
cannot reproduce de Vaucouleur's law, neither the $r^{-4}$ density
decrease of these objects. Indeed, for stellar polytropes, the density
decreases as $r^{-5}$ for $n=5$, the density drops to zero at a finite
distance for $n<5$ and the mass is infinite for $n>5$. We conclude,
therefore, that observations of elliptical galaxies do not support the
prediction of Tsallis. The model of Hjorth \& Madsen (1993) based on
the statistical theory of Lynden-Bell improved by a physically
motivated scenario of incomplete relaxation provides a much better
agreement with observations. This equilibrium state cannot be
described by Tsallis distribution which fails to reproduce
simultaneously an isothermal core and an external confinement.  In
fact, the model of Hjorth \& Madsen (1993) can be obtained by
maximizing a H-function with
\begin{eqnarray}
C({f})=\Biggl\lbrace \begin{array}{cc}
{f}\ln {f},   & {f}>{f}_{0}; \\
{1\over q-1}({f}^{q}-{f}),  & {f}<{f}_{0}.
\end{array}
\label{hm2}
\end{eqnarray}  
at fixed mass and energy. Therefore, their model is consistent with a
{\it composite} Tsallis model with $q=1$ (complete mixing) in the core
and $q\equiv (2n-1)/(2n-3)=7/5$ (incomplete mixing) in the halo. More
generally, we could argue that $q$ varies in space in order to take
into account a variable mixing efficiency (very similarly, stars are
sometimes approximated by composite polytropic models with a space
dependent index in order to distinguish between convective and
radiative regions). However, this gives to Tsallis distribution a
practical, not a fundamental justification. It can just {\it fit}
locally the distribution function of collisionless stellar systems
resulting from incomplete violent relaxation.

In addition, not all stable stationary solutions of the Vlasov equation can be
obtained by maximizing a $H$-function at fixed mass and energy. This
maximization problem leads to $f=f(\epsilon)$ with
$f'(\epsilon)<0$. For spherical stellar systems, the
distribution function can depend on the specific angular momentum
${\bf j}={\bf r}\times {\bf v}$ in addition to the stellar energy
$\epsilon$ and this is what happens in certain models of incomplete
violent relaxation with anisotropic velocity distribution (Lynden-Bell
1967). On the other hand, for rotating stellar systems, the
maximization of $S[{f}]$ at fixed mass, energy and angular momentum
predicts that $f=f(\epsilon_{J})$ with $f'<0$, where $\epsilon_{J}$ is
the Jacobi energy. However, it is well-known that such distribution
function cannot account for the triaxial structure of elliptical
galaxies. More general distribution functions must be constructed in
agreement with the Jeans theorem (Binney \& Tremaine 1987). However,
it is possible that relevant distribution functions maximize a
$H$-function with additional constraints. These constraints may
correspond to adiabatic invariants that are approximately conserved
during violent relaxation. For example, anisotropic velocity
distributions can be obtained by maximizing a $H$-function while
conserving the individual distributions of angular momentum ${\bf
L}_{2}=\int f ({\bf r}\times{\bf v})^{2}d^{3}{\bf r}d^{3}{\bf v}$ (or
other distribution) in addition to mass and energy. This leads to
$f=f(\epsilon_{a})$ with $\epsilon_{a}\equiv
\epsilon+{j^{2}\over 2r_{a}}$ and $f'(\epsilon_{a})<0$ ($r_{a}$ is
called the anisotropy radius). It is expected that this maximization
problem is a condition of nonlinear stability. For rotating systems, it
is not known whether we can construct distribution functions yielding
triaxial bodies by maximizing a $H$-function at fixed $M$, $E$, ${\bf
L}$ and additional (adiabatic) invariants. These ideas may be a route
to explore.

\subsection{Dynamical stability of collisionless stellar systems}
\label{sec_dyns}

On a formal point of view, the maximization problem determining the
nonlinear stability of a collisionless stellar system is {\it similar}
to the one which determines the thermodynamical stability of a
collisional stellar system. In thermodynamics, we maximize the
Boltzmann entropy $S_{B}[f]$ at fixed mass $M$ and energy $E$ in order
to determine the most probable equilibrium state. In stellar dynamics,
the maximization of a H-function $S[f]$ at fixed mass and energy
determines a nonlinearly stable stationary solution of the
Vlasov-Poisson system. Therefore, we can use a {\it thermodynamical
analogy} to analyse the dynamical stability of collisionless stellar
systems. In this analogy, the functional $S[f]$ plays the role of a
generalized entropy, the Lagrange multiplier $\beta$ plays the role of
a generalized inverse temperature $\beta=1/T$, the curve $\beta(E)$
the role of a generalized caloric curve and the maximization of $S[f]$
at fixed $E$ and $M$ corresponds to a microcanonicial description.
This analogy explains why we take $C$ to be convex (and not concave)
in Eq. (\ref{st0}). Then, we must look for maxima of $S$ (not minima)
like in thermodynamics. We can also introduce a $J$-function defined
by $J[f]\equiv S-\beta E$ which is similar to a free energy in
thermodynamics ($J$ is the Legendre transform of $S$). The
maximization of $J[f]$ at fixed $M$ and $\beta$ corresponds to a
canonical description.

The condition that  $f$ is a {maximum} of 
$S$ at fixed mass and energy is equivalent to the
condition that $\delta^{2}{J}=
\delta^{2}{S}-\beta\delta^{2}E$ is negative for all perturbations that
conserve mass and energy to first order. This condition can be written
\begin{eqnarray}
\label{ds1}
\delta^{2}J=-\int C''(f){(\delta f)^{2}\over 2}d^{3}{\bf  r} d^{3}{\bf  v}-{1\over 2}\beta\int \delta\rho\delta\Phi d^{3}{\bf  r}\le 0,\nonumber\\
\forall\ \delta f \mid\ \delta E=\delta M=0.\qquad\qquad \qquad
\end{eqnarray}
The condition that $f$ is a {maximum} of 
$J=S-\beta E$ at fixed mass and temperature is equivalent to the condition that $\delta^{2}{J}$ is negative for all perturbations that
conserve mass. This can be written
\begin{eqnarray}
\label{ds2}
\delta^{2}J=-\int C''(f){(\delta f)^{2}\over 2}d^{3}{\bf  r} d^{3}{\bf  v}-{1\over 2}\beta\int \delta\rho\delta\Phi d^{3}{\bf  r}\le 0,\nonumber\\
\forall\ \delta f \mid\ \delta M=0.\qquad\qquad \qquad
\end{eqnarray} 
According to Eq. (\ref{st2a}), $\delta^{2}J$ can be re-written
\begin{eqnarray}
\label{ds2new}
\delta^{2}J=-{1\over 2}\beta\biggl \lbrace \int {(\delta f)^{2}\over -f'(\epsilon)}d^{3}{\bf  r} d^{3}{\bf  v}+\int \delta\rho\delta\Phi d^{3}{\bf  r}\biggr\rbrace ,
\end{eqnarray}
which is the usual form considered in the literature. We emphasize,
however, that the role of the first order constraints is crucial in
the stability analysis and is often underestimated.  If $f$ is a
maximum of $J$ at fixed $M$ and $\beta$ then it is necessarily a
maximum of $S$ at fixed $M$ and $E$. Indeed, if the inequality
(\ref{ds2}) is satisfied for all perturbations that conserve mass, it
is a fortiori satisfied for perturbations that conserve mass {\it and}
energy. However, the reciprocal is wrong in general. In
thermodynamics, this means that canonical stability implies
microcanonical stability (but the converse is wrong in
general). Exploiting this thermodynamical analogy, we can use the
turning point arguments of Katz (1978) to settle whether a solution of
the form (\ref{st1}) is a maximum or a minimum (or a saddle point) of
$S$ or $J$. The canonical and microcanonical descriptions will be
inequivalent if the curve $\beta(E)$ presents turning points leading
to regions where $C=dE/dT<0$. This corresponds to regions of negative
specific heats in thermodynamics. In these regions, the curve $S(E)$
has a convex dip. This convex dip is the signal of phase transitions
in thermodynamics.

We can also try to reduce the stability problem to the study of an
eigenvalue equation. To that purpose, it is convenient to first
maximize $S[f]$ at fixed mass, energy {\it and} density $\rho({\bf
r})$ in order to obtain a functional $S[\rho]$. Now, maximizing $S[f]$
at fixed mass, energy and density is equivalent to maximizing $S[f]$
at fixed kinetic energy $K$ and density.  Introducing a Lagrange
multiplier $\beta$ for the kinetic energy and a space dependant
Lagrange multiplier $\lambda({\bf r})$ for the density constraint, we
get
\begin{eqnarray}
f=F\biggl\lbrack \beta\biggl ({v^{2}\over 2}+\lambda({\bf r})\biggr )\biggr\rbrack,
\label{st5}
\end{eqnarray}
where $F$ is the same function as in Sec. \ref{sec_vp}. Since
\begin{eqnarray}
\delta^{2}{S}=-\int C''(f){(\delta f)^{2}\over 2}d^{3}{\bf r}d^{3}{\bf v}\le 0,
\label{st6}
\end{eqnarray}
we conclude that Eq. (\ref{st5}) is a {\it maximum } of ${S}$ at fixed
$K$ and $\rho$. This is also a maximum of ${J}={S}-\beta E$ at fixed
$\beta$ and $\rho$.

With the distribution function (\ref{st5}), we can write the density and the pressure as
\begin{eqnarray}
\rho={1\over \beta^{3/2}}g(\beta\lambda), \qquad p={1\over \beta^{5/2}}h(\beta\lambda),
\label{st7}
\end{eqnarray}
where
\begin{eqnarray}
g(x)=\int_{0}^{+\infty}4\pi\sqrt{2}F(t+x)t^{1/2}dt,
\label{st8}
\end{eqnarray}
and
\begin{eqnarray}
h(x)=\int_{0}^{+\infty}{8\pi\sqrt{2}\over 3}F(t+x)t^{3/2}dt.
\label{st9}
\end{eqnarray}
We can now express $E$ and $S$ as functionals of $\rho$ and $\beta$. The energy (\ref{ens2}) is simply given by
\begin{eqnarray}
E={3\over 2}\int p d^{3}{\bf r}+{1\over 2}\int \rho\Phi d^{3}{\bf r}.
\label{st10}
\end{eqnarray}
On the other hand, according to Eq. (\ref{st0}), we have
\begin{eqnarray}
{S}=-{4\pi\sqrt{2}}\int d^{3}{\bf r}\int_{\lambda}^{+\infty}C\lbrack F(\beta\epsilon)\rbrack (\epsilon-\lambda)^{1/2}d\epsilon.
\label{st11}
\end{eqnarray}
Setting $t=\beta(\epsilon-\lambda)$, we get
\begin{eqnarray}
{S}=-{4\pi\sqrt{2}\over\beta^{3/2}}\int d^{3}{\bf r}\int_{0}^{+\infty}C\lbrack F(t+\beta\lambda)\rbrack \ t^{1/2}dt.
\label{st12}
\end{eqnarray}
Integrating by parts  and using $C'\lbrack F(x)\rbrack=-x$, we find that
\begin{eqnarray}
{S}=-{8\pi\sqrt{2}\over 3\beta^{3/2}}\int d^{3}{\bf r}\int_{0}^{+\infty}
F'(t+\beta\lambda)(t+\beta\lambda)t^{3/2}dt.
\label{st13}
\end{eqnarray}
Integrating by parts one more time and using Eqs. (\ref{st7}), (\ref{st8}) and (\ref{st9}), we finally obtain
\begin{eqnarray}
{S}={5\over 2}\beta\int p d^{3}{\bf r}+\beta\int \lambda\rho d^{3}{\bf r}.
\label{st14}
\end{eqnarray}

We now consider the maximization of ${S}[\rho]$ at fixed mass and
energy. This is similar to a microcanonical situation in
thermodynamics. In that case, $\beta$ is not fixed and its variation
$\delta\beta$ is related to the change of density $\delta\rho$ via the
energy constraint (\ref{st10}), using Eqs. (\ref{st7}), (\ref{st8})
and (\ref{st9}). This implies in particular that the barotropic relation
$p=p(\rho)$, valid at equilibrium, is {\it not} preserved when we
consider variations around equilibrium. Indeed, eliminating
$\beta\lambda$ in Eq. (\ref{st7}), we find that
$p=\beta^{-5/2}H(\beta^{3/2}\rho)$ so that $\beta$ enters explicitly
in the relation between $p$ and $\rho$. In the case of isothermal
stellar systems and stellar polytropes described by Boltzmann and
Tsallis entropies, it is possible to reduce the stability problem to
the study of an eigenvalue equation (Padmanabhan 1989, Taruya \&
Sakagami 2002a). However, it seem difficult to make this reduction in the
general case.

\subsection{Dynamical stability of gaseous spheres}
\label{sec_dsgs}

We now consider the maximization of $J[\rho]=S[\rho]-\beta E[\rho]$ at
fixed $\beta$ and $M$. This is similar to a canonical situation in
thermodynamics. According to Eqs. (\ref{st10}) and (\ref{st14}), we
have
\begin{eqnarray}
{J}=\beta\int p d^{3}{\bf r}+\beta\int \lambda\rho d^{3}{\bf r}-{1\over 2}\beta\int \rho\Phi d^{3}{\bf r},
\label{st15}
\end{eqnarray}
Since $\beta$ is constant, the barotropic relation $p=p(\rho)$ remains
valid around equilibrium.  Using the relation $h'(x)=-g(x)$ obtained 
from Eqs. (\ref{st8}) and (\ref{st9}) by a simple integration
by parts, it is easy to check that Eq. (\ref{st7}) yields
\begin{eqnarray}
\lambda+{p\over\rho}=-\int_{0}^{\rho}{p(\rho')\over\rho'^{2}}d\rho'.
\label{st16}
\end{eqnarray} 
Therefore, the functional (\ref{st15}) can be
rewritten as
\begin{eqnarray}
{J}=-\beta\int \rho \int_{0}^{\rho}{p(\rho')\over\rho'^{2}}d\rho' d^{3}{\bf r}-{1\over 2}\beta\int \rho\Phi d^{3}{\bf r}.
\label{st17}
\end{eqnarray}
If we cancel the first order variations of ${J}$ at fixed mass and $\beta$, we get $\lambda=\Phi+\alpha/\beta$ and we recover Eq. (\ref{st2}). On the other hand, the condition that  Eq. (\ref{st2}) is a maximum of $J$ at fixed $M$ and $\beta$  requires that the second order variations of ${J}$ be negative. This leads to the inequality
\begin{eqnarray}
\delta^{2}{J}=-\beta\int {p'(\rho)\over 2\rho} (\delta\rho)^{2} d^{3}{\bf r}-{1\over 2}\beta\int \delta\rho\delta\Phi d^{3}{\bf r}\le 0,
\label{st18}
\end{eqnarray}
for all perturbations $\delta\rho$ that conserve mass. Using the condition of hydrostatic equilibrium, $\delta^{2}J$ can be re-written
\begin{eqnarray}
\delta^{2}{J}=-{1\over 2}\beta\biggl\lbrace \int {(\delta\rho)^{2}\over -\rho'(\Phi)} d^{3}{\bf r}+\int \delta\rho\delta\Phi d^{3}{\bf r}\biggr\rbrace.
\label{st18new}
\end{eqnarray}
We note, in passing, the similarity with the pseudo-energy (or Arnol'd
invariant) in two-dimensional hydrodynamics (see Chavanis
2002h). Therefore, Arnol'd theorems can be directly extended to (3D)
self-gravitating bodies. In particular, in a bounded domain, the
system is nonlinearly stable, for any type of perturbation, if $0\le
-\rho'(\Phi)\le {\lambda_{1}/(4\pi G)}$, where $\lambda_{1}$ is the
smallest eigenvalue of $-\Delta$ (for isothermal spheres this
corresponds to $\alpha\le \pi$ and for polytropic spheres to
$\alpha\le \pi/\sqrt{n}$ where $\alpha$ is defined by Eqs. (\ref{alp})
and (\ref{lane3}) respectively).  Furthermore, we have already shown
in the case of isothermal stellar systems and stellar polytropes that
the stability condition (\ref{st18}) can be reduced to the study of an
eigenvalue equation. This reduction can be generalized to any stellar
system described by a H-function $S[f]$ of the form
(\ref{st0}). Considering spherically symmetrical perturbations,
introducing the variable $q(r)$ defined by Eq. (\ref{TsallisC3}) and
repeating exactly the same steps as for isothermal and polytropic
spheres (see Chavanis 2002a and Sec. \ref{sec_TsallisC}), we end up on
the eigenvalue problem
\begin{eqnarray}
\biggl\lbrack {d\over dr}\biggl ({p'(\rho)\over 4\pi \rho r^{2}}{d\over dr}\biggr )+{G\over r^{2}}\biggr\rbrack q_{\lambda}(r)=\lambda q_{\lambda}(r),
\label{st19}
\end{eqnarray}
with appropritate boundary conditions (see Appendix \ref{sec_puls}).
If all the $\lambda$'s are negative, then  Eq. (\ref{st2}) is a maximum of 
$J$ at fixed $M$ and $\beta$.

We now consider the stability of a gaseous system with an equation of
state $p=p(\rho)$. The energy of this gaseous configuration is
(Lynden-Bell \& Sanitt 1969)
\begin{eqnarray}
{\cal W}=\int \rho \int_{0}^{\rho}{p(\rho')\over\rho'^{2}}d\rho' d^{3}{\bf r}+{1\over 2}\int \rho\Phi d^{3}{\bf r}.
\label{st17bis}
\end{eqnarray}
We observe that the $J$-function (\ref{st17}) associated with a
stellar system is equal (up to a negative proportionality factor
$-\beta$) to the energy (\ref{st17bis}) of the corresponding
barotropic gas.  The stability theorem of Chandrasekhar for gaseous
spheres requires that $\delta^{2}{\cal W}\ge 0$ (see Binney \&
Tremaine 1987). This leads to the inequality (\ref{st18}) and to the
eigenvalue equation (\ref{st19}). We now consider the dynamical
stability of a self-gravitating gaseous system with respect to the 
Navier-Stokes or Euler equations (Jeans problem). This problem can be
solved explicitly for isothermal and polytropic spheres (Chavanis
2002a,b). Adapting the same procedure for a general barotropic
equation of state and denoting by $\lambda$ the growth rate of the
perturbation (such that $\delta\rho\sim e^{\lambda t}$ etc...), we
obtain the eigenvalue equation
\begin{eqnarray}
\biggl\lbrack {d\over dr}\biggl ({p'(\rho)\over 4\pi \rho r^{2}}{d\over dr}\biggr )+{G\over r^{2}}\biggr\rbrack q_{\lambda}(r)={\lambda^{2}\over 4\pi\rho r^{2}} q_{\lambda}(r),
\label{st20}
\end{eqnarray}
with appropriate boundary conditions (see Appendix \ref{sec_puls}).
This equation is equivalent to the Eddington (1926) equation of
pulsation but it proved to be more convenient in our previous
investigations. We observe that Eqs. (\ref{st19}) and (\ref{st20}) are
similar and that they coincide for marginal stability. This first
demonstates the equivalence between the stability criterion of
Eddington based on the equation of pulsation (\ref{st20}) and the
stability criterion of Chandrasekhar based on the minimization of the
energy (\ref{st17bis}). This also shows the {\it equivalence} between
the maximization of the $J$-function for a stellar system at fixed
$\beta$ and $M$ and the dynamical stability (with respect to the
Jeans-Euler equations) of the corresponding barotropic star. Note
that we have only studied the dynamical stability of barotropic stars
with respect to small perturbations. However, we conjecture that the
maximization of $J$ at fixed $M$ and $\beta$ is a condition of {\it
nonlinear} dynamical stability for the Jeans-Euler equation. The
proof is expected to be similar to the corresponding one for the
Vlasov equation.

\subsection{A new interpretation of Antonov's first law}
\label{sec_anto}

We can now combine the previous results in order to provide a new
interpretation of Antonov's first law (Binney \& Tremaine 1987) based
on a thermodynamical analogy. We have seen in Sec. \ref{sec_dyns} that
a stellar system with $f=f(\epsilon)$ and $f'(\epsilon)<0$ is
dynamically stable (nonlinearly) with respect to collisionless
perturbations if $f$ is a maximum of a $H$-function $S[f]$ at fixed
$E$ and $M$. According to the discussion following Eqs. (\ref{ds1})
and (\ref{ds2}), a {\it sufficient} condition of dynamical stability
is that $f$ is a maximum of the $J$-function $J[f]$ at fixed $\beta$
and $M$. Now, according to Sec. \ref{sec_dsgs}, this is equivalent to
the dynamical stability of the corresponding barotropic
gas. Therefore, we come to the conclusion that ``a stellar system is
dynamically stable whenever the corresponding barotropic star is
stable'', which is Antonov's first law. Owing to the thermodynamical
analogy, Antonov's first law is similar to the property that
``canonical stability implies microcanonical stability'' in
thermodynamics. We note that the reciprocal of Antonov's first law is
not true in general. If a stellar system is stable this does not
necessarily imply that the corresponding barotropic gas is
stable. Similarly, in thermodynamics, microcanonical stability does
not necessarily imply canonical stability. This is the case only if
the statistical ensembles are equivalent. We know, however, that this
equivalence is broken for self-gravitating systems.

Let us illustrate these results in the case of polytropes. Taruya \&
Sakagami (2002a,b) have investigated the 
stability of  stellar systems in the framework of Tsallis
generalized thermodynamics (see also Sec. \ref{sec_sp}). Due to the
thermodynamical analogy discussed in Secs. \ref{sec_dyns} and
\ref{sec_dsgs}, the study of Taruya \& Sakagami can be used to
determine the {dynamical stability} of collisionless stellar
polytropes and polytropic gaseous spheres. In the generalized
thermodynamical approach, it is found that the statistical ensembles
are inequivalent. This implies that the dynamical stability of stellar
polytropes and gaseous polytropes will also be inequivalent (Chavanis
2002b). Self-gravitating systems described by Tsallis entropy are
thermodynamically stable in the microcanonical ensemble if $n<5$ or if
the density contrast is sufficiently low for $n>5$. In that case, they
are maxima of $S_{q}[f]$ at fixed $M$ and $E$. According to
Sec. \ref{sec_dyns}, this implies that stellar polytropes are
dynamically stable solutions of the Vlasov equation for $n<5$ (and if
the density contrast is sufficiently low for $n>5$). On the other
hand, self-gravitating systems described by Tsallis entropy are
thermodynamically stable in the canonical ensemble if $n<3$ or if the
density contrast is sufficiently low for $n>3$.  In that case, they
are maxima of $J_{q}[f]$ at fixed $M$ and $\beta$. According to
Sec. \ref{sec_dsgs}, this implies that polytropic gaseous spheres are
dynamically stable solutions of the Jeans-Euler equations if, and
only, if $n<3$ (or if, and only, if the density contrast is
sufficiently low for $n>3$).  We note, in particular, that gaseous
polytropes with $3<n<5$ are unstable for sufficiently high density
contrasts (e.g., complete polytropes) while corresponding stellar
polytropes are stable. This shows that the reciprocal of Antonov's
first law is wrong for polytropes. Due to the thermodynamical analogy,
this can be related to an inequivalence of statistical ensembles in a
region of negative specific heats (Chavanis 2002b). Our present study
shows that the results obtained for isothermal stellar systems and
stellar polytropes can be generalized to any $H$-function.

\section{Conclusion}
\label{sec_conc}

We have performed an exhaustive study of the thermodynamics of
self-gravitating systems in various ensembles. This paper completes
previous investigations on the subject and all ensembles have now been
treated. Contrary to ordinary (extensive) systems, we have to perform
a specific study in each ensemble since the stability limits differ
from one to the other. Remarkably, the thermodynamical stability
problem can be studied analytically or with simple graphical
constructions. The dynamical stability of isothermal gaseous spheres
can also be studied analytically both in Newtonian mechanics for the
Euler-Jeans equations (Chavanis 2002a) and in general relativity for
the Einstein equations (Chavanis 2002d).

These results can be of relevance both for astrophysicists and
statistical mechanicians and could make a bridge between these two
communities. On an astrophysical point of view, they show that we
must be careful to precisely define the ensemble in which we work
since they are not equivalent. This does not affect the structure of
the equilibrium configuration but it may affect its stability. On a
physical point of view, this study fills an important gap in the
statisical mechanics literature since the case of self-gravitating
systems is not discussed at all in standard textbooks of statistical
mechanics and thermodynamics. 

We have also discussed the relevance of Tsallis generalized
thermodynamics for stellar systems. Collisionless stellar systems such
as elliptical galaxies can achieve a metaequilibrium state as a result
of a violent relaxation.  The Boltzmann entropy $S_{B}[\overline{f}]$
is the correct entropy for these systems (in a coarse-grained sense
and assuming that the system is non-degenerate). It measures the
disorder where the disorder is equal to the number of microstates
consistent with a given macrostate. However, $S_{B}[\overline{f}]$ is
not maximized by the system because of incomplete relaxation (this is
independant on the fact that $S_{B}[\overline{f}]$ has no
maximum!). In any case, the state resulting from incomplete violent
relaxation is a nonlinearly stable solution of the Vlasov equation (on
a coarse-grained scale). If $\overline{f}=f(\epsilon)$, it maximizes a
H-function $S[f]=-\int C(f)d^{3}{\bf r}d^{3}{\bf v}$, where $C(f)$ is
a convex function, at fixed mass and energy. Therefore, we can use a
thermodynamical analogy to study the dynamical stability of
collisionless stellar systems. Tsallis entropies $S_{q}[f]$ are a
particular class of H-functions leading to stellar polytropes. Stellar
polytropes do not give a good fit of elliptical galaxies. A better
model of incomplete violent relaxation, motivated by precise physical
arguments, consists of an isothermal core and a polytropic halo with
index $n\simeq 4$ (Hjorth
\& Madsen 1993). This can be fitted by  a composite 
Tsallis (polytropic) model with $q=1$ in the core (complete mixing)
and $q\simeq 7/5$ in the envelope (incomplete mixing). More generally,
the state resulting from incomplete violent relaxation does not
necessarily maximize a H-function at fixed mass and energy. Due to the
Jeans theorem, the distribution function can depend on other integrals
than the stellar energy or the Jacobi energy. It is possible,
nevertheless, that relevant distribution functions arising from
incomplete violent relaxation maximize a $H$-function at fixed mass,
energy, angular momentum and additional (adiabatic)
invariants. 

\vskip0.2cm
\noindent{\it Acknowledgements.} I am grateful to J. Katz and T. Padmanabhan
for stimulating discussions and encouragement.

\appendix

\section{Some useful identities for isothermal spheres}
\label{sec_id1}

Using the Emden equation (\ref{emden2}), we have
\begin{eqnarray}
\int_{0}^{\alpha}\xi^{2}e^{-\psi}d\xi=\int_{0}^{\alpha} (\xi^{2}\psi')'d\xi=\alpha^{2}\psi'(\alpha)=\alpha v_{0},
\label{id1}
\end{eqnarray}
which establishes Eq. (\ref{gmc20}). On the other hand, using an integration by
parts, we obtain
\begin{eqnarray}
\int_{0}^{\alpha}\psi'\xi^{3}e^{-\psi}d\xi=-\int_{0}^{\alpha} \xi^{3}{d\over d\xi}(e^{-\psi})d\xi\nonumber\\
=-\alpha^{3}e^{-\psi(\alpha)}+\int_{0}^{\alpha}3\xi^{2}e^{-\psi}d\xi.
\label{id2}
\end{eqnarray}
Combining with Eq. (\ref{id1}) and introducing the Milne variables, we establish Eq. (\ref{gmc21}). To establish Eq. (\ref{gmc22}), we start from the identity
\begin{eqnarray}
\int_{0}^{\alpha}{\xi^{2}\psi'\over \xi}{d\over d\xi}(\xi^{2}\psi')d\xi=\alpha^{3}\psi'(\alpha)^{2} \nonumber\\
-\int_{0}^{\alpha}{\xi^{2}\psi'\over \xi}{d\over d\xi}(\xi^{2}\psi')d\xi   
+\int_{0}^{\alpha}\xi^{2}\psi'^{2}d\xi,
\label{id3}
\end{eqnarray}
which results from a simple integration by parts. Thus,
\begin{eqnarray}
\int_{0}^{\alpha}\xi^{2}\psi'^{2}d\xi=-\alpha^{3}\psi'(\alpha)^{2}+2\int_{0}^{\alpha}\psi'\xi^{3}e^{-\psi}d\xi,
\label{id4}
\end{eqnarray}
where we have used the Emden equation (\ref{emden2}). Combining this
relation with Eq. (\ref{id2}), we obtain
\begin{eqnarray}
\int_{0}^{\alpha}\xi^{2}\psi'^{2}d\xi=\alpha v_{0}(6-v_{0}-2u_{0}).
\label{id5}
\end{eqnarray}
Now,
\begin{eqnarray}
\int_{0}^{\alpha}\xi^{2}\psi e^{-\psi}d\xi=\int_{0}^{\alpha}\psi (\xi^{2}\psi')'d\xi\nonumber\\
=\alpha^{2}\psi(\alpha)\psi'(\alpha)-\int_{0}^{\alpha}\xi^{2}
\psi'^{2}d\xi. 
\label{id6}
\end{eqnarray}
From  Eqs. (\ref{id5}) and (\ref{id6}) results Eq. (\ref{gmc22}).

\section{Some useful identities for polytropic spheres}
\label{sec_id2}

Using integrations by parts similar to those performed in Appendix \ref{sec_id1}, we can easily establish the identities
\begin{eqnarray}
\int_{0}^{\alpha}(\theta')^{2}\xi^{2}d\xi={\alpha^{3}\theta'(\alpha)^{2}\over 5-n}\biggl (n+1+2{u_{0}\over v_{0}}-{6\over v_{0}}\biggr ),
\label{idp1}
\end{eqnarray}
\begin{eqnarray}
\int_{0}^{\alpha}\theta^{n+1}\xi^{2}d\xi={\alpha^{3}\theta'(\alpha)^{2}\over 5-n}\biggl (n+1+2{u_{0}\over v_{0}}-{n+1\over v_{0}}\biggr ),
\label{idp2}
\end{eqnarray}
\begin{eqnarray}
\int_{0}^{\alpha}\theta^{n}\theta'\xi^{3}d\xi={\alpha^{3}\theta'(\alpha)^{2}\over 5-n}\biggl (-3-{u_{0}\over v_{0}}+{3\over v_{0}}\biggr ),
\label{idp3}
\end{eqnarray}
\begin{eqnarray}
\int_{0}^{\alpha}\theta^{n}\xi^{2}d\xi=-\alpha^{2}\theta'(\alpha).
\label{idp4}
\end{eqnarray}

\section{Equation of pulsations and boundary conditions}
\label{sec_puls}

In this paper and previous ones, we have studied the dynamical stability of barotropic gaseous spheres with the equation of pulsation
\begin{eqnarray}
\biggl\lbrack {d\over dr}\biggl ({p'(\rho)\over 4\pi \rho r^{2}}{d\over dr}\biggr )+{G\over r^{2}}\biggr\rbrack q_{\lambda}(r)={\lambda^{2}\over 4\pi\rho r^{2}} q_{\lambda}(r),
\label{puls1}
\end{eqnarray}
where $\lambda$ is the growth rate of the perturbation (such that $\delta\rho\sim e^{\lambda t}$, $\delta v\sim e^{\lambda t}$ etc...) and $q(r)\equiv\delta M(r)=\int_{0}^{r}\delta\rho 4\pi r^{2}dr$ represent the mass perturbation within the sphere of radius $r$ which is related to the velocity by (Chavanis 2002a)
\begin{eqnarray}
\delta v=-{\lambda\over 4\pi \rho r^{2}}q.
\label{puls2}
\end{eqnarray}
Obviously $q(0)=0$ and more precisely $q\sim r^{3}$ for $r\rightarrow
0$. If the system is confined within a sphere of fixed radius $R$, the
conservation of mass imposes $q(R)=0$. This can also be obtained from
Eq. (\ref{puls2}) if we set $\delta v(R)=0$. If the system is free and
the density vanishes at its surface, i.e. $\rho(R)=0$, like for a
complete polytrope, then Eq. (\ref{puls2}) again implies that $q(R)=0$
(Chavanis 2002b). If, finally, the gaseous sphere supports a fixed external
pressure (isobaric ensemble) the boundary condition requires that the
Lagrangian derivative of the pressure vanishes at the surface of the
configuration, i.e.
\begin{eqnarray}
{d\over dt}\delta p+\delta v{dp\over dr}=0, \qquad {\rm at}\ r=R.
\label{puls3}
\end{eqnarray}
Using $d\delta p/dt=\lambda \delta p$ and Eqs. (\ref{puls2}) and (\ref{TsallisC3}),
we get for an arbitrary  barotropic equation of state 
\begin{eqnarray}
{dq\over dr}-{q\over \rho}{d\rho\over dr}=0, \qquad {\rm at}\ r=R.
\label{puls4}
\end{eqnarray}

If we set $\delta v=d\delta r/dt=\lambda\delta r$ and introduce the
Lagrangian displacement $\xi(r)\equiv \delta r/r=\delta v/\lambda r$,
we can easily check that Eq. (\ref{puls1}) is equivalent to the
Eddington (1926) equation of pulsations
\begin{eqnarray}
{d\over dr}\biggl (p\gamma r^{4}{d\xi\over dr}\biggr )+\biggl\lbrace r^{3}{d\over dr}\lbrack (3\gamma-4)p\rbrack-\lambda^{2}\rho r^{4}\biggr\rbrace\xi=0,
\label{pulse5}
\end{eqnarray}
where $\gamma(r)\equiv d\ln p/d\ln \rho$. The boundary condition at the origin is $\xi'(0)=0$. On the other hand, if the system is enclosed within a box $\xi(R)=0$. Finally, if the pressure at the surface of the sphere is constant (possibly zero), the boundary condition (\ref{puls4}) becomes  
\begin{eqnarray}
{d\xi\over dr}=-{3\xi\over r}, \qquad {\rm at}\ r=R.
\label{pulse6}
\end{eqnarray}


\begin{thebibliography}{}

\bibitem{antonov}  {\small Antonov, V.A. 1962, Vest. Leningr. Gos. Univ., {7}, 135}

\bibitem{aronson}  {\small Aronson, E.B. \& Hansen, C.J. 1972, ApJ, {177}, 145}

\bibitem{bettwieser}  {\small Bettwieser, E. \& Sugimoto, D. 1984, MNRAS, {208}, 493}

\bibitem{bilic}  {\small Bilic, N. \& Viollier, R.D. 1997, Phys. Lett. B, {408}, 75}

\bibitem{bt}  {\small  Binney, J., Tremaine, S. 1987,
Galactic Dynamics (Princeton Series in Astrophysics)}

\bibitem{bonnor}  {\small Bonnor, W.B.  1956, MNRAS, {116}, 351}

\bibitem{chandra}  {\small  Chandrasekhar, S. 1942, 
An introduction to the theory of stellar Structure (Dover)}  

\bibitem{trunc}  {\small Chavanis, P.H. 1998, MNRAS, 300, 981}

\bibitem{chaviso}  {\small Chavanis, P.H. 2002a, A\&A, 381, 340 }

\bibitem{chavpoly}  {\small Chavanis, P.H. 2002b, A\&A, 386, 732 }

\bibitem{chavtrans}  {\small Chavanis, P.H. 2002c, PRE,  65, 056123 }

\bibitem{chavrelat}  {\small Chavanis, P.H. 2002d, A\&A, 381, 709}

\bibitem{dauxois}  {\small  Chavanis, P.H. 2002e, Statistical mechanics of two-dimensional vortices and stellar systems, in: Dynamics and thermodynamics of systems with long range interactions, Eds: Dauxois, T, Ruffo, S., Arimondo, E. and  Wilkens, M. Lecture Notes in Physics, vol 602, Springer; e-print cond-mat/0212223}

\bibitem{dubro}  {\small  Chavanis, P.H. 2002f, Statistical mechanics of violent relaxation in stellar systems, in: Proc. of the Conference on Multiscale Problems in Science and Technology (Springer); e-print astro-ph/0212205}

\bibitem{cape}  {\small  Chavanis, P.H. 2002g, The self-gravitating Fermi gas, in:  Proceedings of the Fourth International Heidelberg Conference on Dark Matter in Astro and Particle Physics, Eds: Klapdor-Kleingrothaus  (Springer, New-York, 2002); e-print astro-ph/0205426} 

\bibitem{gt}  {\small Chavanis, P.H. 2002h, submitted to PRE 
[cond-mat/0209096] }

\bibitem{ispo}  {\small Chavanis, P.H. \& Ispolatov, I. 2002, PRE, 66, 036109 }

\bibitem{cso}  {\small  Chavanis, P.H., Sommeria, J.  1998, 
MNRAS, {296}, 569}

\bibitem{csr}  {\small  Chavanis, P.H., Sommeria, J. \& Robert, R.  1996, 
ApJ, {471}, 385}

\bibitem{crs}  {\small  Chavanis, P.H., Rosier, C. \& Sire, C. 2002 , PRE, 66, 036105}

\bibitem{cohn}  {\small  Cohn, H. 1980, 
ApJ, {242}, 765}

\bibitem{deVega}  {\small de Vega, H.J., Sanchez, N. \& Combes, F.  1998, ApJ,
 500, 8}

\bibitem{deVega2}  {\small de Vega, H.J., Sanchez, N. 2002, Nucl. Phys. B, 625, 409}

\bibitem{doremus}  {\small Doremus, J.P., Feix, M.R. \& Baumann, G.  1971, Phys. Rev. Lett., 26, 725}

\bibitem{ebert}  {\small Ebert, R. 1955, Z. Astrophys., 37, 217}

\bibitem{eddington}  {\small  Eddington, A.S. 1926 {The internal constitution of stars}, Cambridge University Press}

\bibitem{hawking}  {\small Hawking, S.W.  1974, Nature, 248, 30}

\bibitem{henon}  {\small H\'enon, M. 1961, Ann. Astrophys. {24}, 369}

\bibitem{hertel}  {\small  Hertel, P. \& Thirring, W. 1971, in: Quanten und Felder, edited by H.P. D\"urr (Vieweg, Braunschweig)} 

\bibitem{hjorth}  {\small Hjorth, J. \& Madsen, J.  1993,   MNRAS {265}, 237}

\bibitem{horwitz}  {\small Horwitz, G. \&  Katz, J. 1978, ApJ, 222, 941}

\bibitem{inagaki}  {\small Inagaki, S. \& Lynden-Bell, D.  1983, MNRAS, 205, 913}

\bibitem{ipser}  {\small Ipser, J.R. 1974, ApJ, 193, 463}

\bibitem{ipser2}  {\small Ipser, J.R. \&  Horwitz, G. 1979, ApJ, 232, 863}

\bibitem{jaffe}  {\small Jaffe, W.  1987, in: Structure and dynamics of elliptical galaxies, Proc. IAU Symp, 127,  Eds: T. de Zeeuw (Reidel, Dordrecht)}

\bibitem{kandrup}  {\small Kandrup, H.E. 1981, ApJ, 244, 316}

\bibitem{katz}  {\small Katz, J. 1978,   MNRAS {183}, 765}

\bibitem{katz2}  {\small Katz, J. 1980,   MNRAS {190}, 497}

\bibitem{okamoto}  {\small Katz, J. \& Okamoto, I. 2000,   MNRAS {371}, 163}

\bibitem{kiessling}  {\small  Kiessling, M. 1989, J. Stat. Mech.   {55}, 203}

\bibitem{lecar}  {\small  Lecar, M. \& Katz, J. 1981, ApJ  {243}, 983}

\bibitem{lee}  {\small Lee, E.P. 1968, ApJ, 151, 687}

\bibitem{lb}  {\small  Lynden-Bell, D. 1967, MNRAS  {136}, 101}

\bibitem{lbe}  {\small  Lynden-Bell, D. \& Eggleton, P.P. 1980, MNRAS  {191}, 483}

\bibitem{lbs}  {\small  Lynden-Bell, D. \& Sanitt, N. 1969, MNRAS  {143}, 167}

\bibitem{lbw}  {\small  Lynden-Bell, D. \& Wood, R. 1968,  
MNRAS, {138}, 495 }

\bibitem{McCrea}  {\small  McCrea, W.H. 1957,  
MNRAS, {117}, 562 }

\bibitem{Michel}  {\small  Michel, J. \& Robert, R. 1994,  
Commun. Math. Phys., {159}, 195 }

\bibitem{ogorodnikov}  {\small  Ogorodnikov, K.F. 1965, Dynamics of stellar systems (Pergamon) }

\bibitem{pad2}  {\small  Padmanabhan, T. 1989,  ApJ Supp.,  {71}, 651 }

\bibitem{pad}  {\small Padmanabhan, T. 1990, Phys. Rep.,  {188}, 285 }

\bibitem{pad3}  {\small Padmanabhan, T. 1991, MNRAS,  {253}, 445}

\bibitem{penston}  {\small Penston, M.V. 1969, MNRAS,  {144}, 425}

\bibitem{plastino}  {\small Plastino, A. \& Plastino, A.R. 1997, Phys. Lett. A,  {226}, 257 }

\bibitem{plummer}  {\small Plummer, H.C.  1911, MNRAS,  {71}, 460 }

\bibitem{sc}  {\small  Sire, C. \& Chavanis, P.H.  2002, PRE, 66, 046133}

\bibitem{taruya1}  {\small Taruya, A. \& Sakagami, M. 2002a,  Physica A, 307, 185} 

\bibitem{taruya2}  {\small Taruya, A. \& Sakagami, M. 2002b,  submitted to Physica A, [cond-mat/0204315]}

\bibitem{terletsky}  {\small Terletsky, Y.P.  1952, Zh. Eksper. Teor. Fiz., 22, 506}

\bibitem{tremaine}  {\small Tremaine, S., H\'enon, M. \& Lynden-Bell, D. 1986, MNRAS,  {227}, 543}

\bibitem{tsallis}  {\small Tsallis, C. 1988, J. Stat. Phys.,  {52}, 479} 

\bibitem{yabushita}  {\small Yabushita, S. 1968, MNRAS,  {140}, 109} 



\end{thebibliography}
\end{document}